\documentclass{myelsart}
\usepackage{amssymb}
\usepackage{booktabs}
\usepackage{hepnicenames,graphicx,colordvi,gnuplotcolors}
\usepackage[dvips]{color}
\usepackage{rotating}

\newcounter{bla}

\input paperdef
\newcommand{\mystar}{$\,^{\star}$}
\newcommand{\mystarbrac}{$\,^{(\star)}$}

\newcommand{\citehdecay}{\cite{hep-ph/9704448}}
\newcommand{\citeSMVHCS}{\cite{hep-ph/0307206,hep-ph/0306234,hep-ph/0406152}}
\newcommand{\citeSMgluonfusionCS}{\cite{PRLTA.70.1372,hep-ph/9504378,hep-ph/0201206,hep-ph/0302135,hep-ph/0207004,hep-ph/0306211,hep-ph/0508265,hep-ph/0508284}}
\newcommand{\citeSMVBFCS}{\cite{hep-ph/0306109,hep-ph/9206246,hep-ph/9905386,hep-ph/0403194}}
\newcommand{\citeSMttHCS}{\cite{hep-ph/0107081,hep-ph/0107101,hep-ph/0211438}}
\newcommand{\feynhiggs}{{\tt FeynHiggs}~\cite{hep-ph/9812320,hep-ph/9812472,hep-ph/0212020,hep-ph/0611326}} 
\newcommand{\cpsuperh}{{\tt CPsuperH}~\cite{hep-ph/0307377,arXiv:0712.2360}} 
\newcommand{\darksusy}{{\tt DarkSUSY}~\cite{astro-ph/0406204}}
\newcommand{\fittino}{{\tt Fittino}~\cite{hep-ph/0412012}}
\newcommand{\mastercode}{{\tt MasterCode}~\cite{arXiv:0707.3447,arXiv:0808.4128}}
\newcommand{\THDMC}{{\tt 2HDMC}~\cite{arXiv:0902.0851}}
\newcommand{\nHplus}{n_{H^{\pm}}}
\newcommand{\nHneut}{n_{H^0}}

\definecolor{gnuplotorange}{rgb}{1,0.647059,0}
\definecolor{gnuplotyellow}{rgb}{1,1,0}
\definecolor{gnuplotred}{rgb}{1,0,0}
\definecolor{gnuplotgreen}{rgb}{0,1,0}
\definecolor{gnuplotdarkblue}{rgb}{0,0,0.545098}
\definecolor{gnuplotlightblue}{rgb}{0.678431, 0.847059, 0.901961}
\definecolor{gnuplotblue}{rgb}{0,0,1}
\definecolor{myblue}{rgb}{0,0,1}

\begin{document}
\begin{frontmatter}

\begin{flushright}
FR-PHENO-2011-002\\
BONN-TH-2011-02\\
DESY 11-016\\
\end{flushright}

\title{{\tt {HiggsBounds} \vers{2}{0}{0}}: 
Confronting Neutral and Charged Higgs Sector Predictions\\ 
with Exclusion Bounds\\
from LEP and the Tevatron
}

\author[DESY]{P.~Bechtle},
\author[FREIBURG]{O.~Brein},
\author[SANTANDER]{S.~Heinemeyer},
\author[DESY]{G.~Weiglein}\\
and
\author[BONN]{K.E.~Williams}.

\address[DESY]{DESY, Notkestrasse 85, 22607 Hamburg, Germany}
\address[FREIBURG]{
Physikalisches Institut,
Albert-Ludwigs-Universit\"at Freiburg,\\
Hermann-Herder-Str. 3, D-79104 Freiburg im Breisgau, Germany} 
\address[SANTANDER]{Instituto de F\'isica de Cantabria (CSIC-UC), 
Santander, Spain}
\address[BONN]{Bethe Center for Theoretical Physics, 
Physikalisches Institut der
Universit\"at Bonn,
Nussallee 12, 53115 Bonn, Germany}

\begin{abstract}
{\tt HiggsBounds} \vers{2}{0}{0}\ is a computer code which tests both neutral and
charged Higgs sectors of arbitrary models against the current exclusion bounds
from the Higgs searches at LEP and the Tevatron.  As input, it requires a
selection of model predictions, such as Higgs masses, branching ratios,
effective couplings and total decay widths. {\tt HiggsBounds} \vers{2}{0}{0} then
uses the expected and observed topological cross section limits from the
Higgs searches to determine 
whether a given parameter scenario of a model is 
excluded at the 95\% C.L. by those searches.
Version 2.0.0 represents a significant extension
of the code since its first release (1.0.0).
It includes now 28/53 LEP/Tevatron Higgs search analyses, 
compared to the 11/22 in the first release,
of which many of the ones from the Tevatron are replaced by updates.
As a major extension, the code allows now the predictions 
for (singly) charged Higgs bosons to be 
confronted with LEP and Tevatron searches.
Furthermore,
the newly included analyses contain
LEP searches for neutral Higgs bosons ($H$) decaying invisibly or into 
(non flavour tagged) hadrons as well as decay-mode independent searches 
for neutral Higgs bosons,
LEP searches via the production modes $\tau^+\tau^- H$ and $b\bar b H$,
and Tevatron searches via $t\bar t H$.
Also, all 
Tevatron results presented at the 
ICHEP'10 are included in version 2.0.0.
As physics applications of
\HB\ \vers{2}{0}{0} we
study the allowed Higgs mass range for model scenarios 
with invisible Higgs decays
and 
we obtain exclusion results for the scalar sector of the Randall-Sundrum
model using up-to-date LEP and Tevatron direct search results.
\end{abstract}
\vspace*{-5mm}
\end{frontmatter}

\section{Introduction}

The search for Higgs bosons is a major cornerstone of the physics programmes
of past, present and future high energy colliders. The LEP and Tevatron
experiments, in particular, have been able to turn the non-observation of
Higgs bosons into constraints on the Higgs sector, which can be very useful
in 
restricting
the available parameter space of particle physics models. 
At the LHC, discoveries involving colour charged
particles (e.g. squarks, gluinos) may occur already before 
a Higgs signal gets established.
The first LHC Higgs search results from ATLAS and CMS are 
expected to become public during the next months. It is important to
note that limits from the Higgs searches at the LHC, the Tevatron and
LEP will play a crucial role also for testing the validity of model
interpretations of possible Higgs-like signals or other signals of new
physics.

The Higgs searches performed by the experimental collaborations
usually take one of two forms. Dedicated analyses have been
carried out in order to constrain some of the most popular models, such as
the Standard Model (SM)~\cite{hep-ex/0306033} and 
various benchmark scenarios in the 
Minimal Supersymmetric Standard Model (MSSM)~\cite{hep-ex/0602042}. 
In addition, (essentially)
model-independent limits on the cross sections of individual
 signal topologies (such as $e^+e^-\to h_iZ\to b\bar{b}Z$) have been
 published. The former type of analyses include detailed knowledge of the
 overlap between the
individual experimental searches, and therefore have a high sensitivity,
whereas the latter can be used to test a wide class of models.

There are certain issues involved with the application of these experimental
constraints. The data is distributed over many different publications and
the limits are given with a variety of normalisations. In the case of the
Tevatron, the results are also frequently updated. Furthermore, care must be
taken when using more than one experimental analysis to ensure the
correct statistical interpretation of the result in terms of a
confidence level (C.L.). One needs to take into account the underlying
assumptions of the various experimental analyses in order to determine 
which analysis is applicable to which Higgs boson of a particular model.
For example, if an analysis has been carried out
assuming the validity of the SM,
it will not be applicable to Higgs bosons
whose various production and decay mechanisms do not show a
proportionate similarity to the SM.

The Fortran code {\tt HiggsBounds} has been designed to facilitate the task
of comparing Higgs sector predictions with existing exclusion limits, thus
allowing the user to quickly and conveniently check a wide variety of models
against the state-of-the-art results from Higgs searches. {\tt HiggsBounds
2.0.0} represents a significant 
extension of
the code since its first release in February 2009
\cite{arXiv:0811.4169} and its latest release in 
December 2009 (version 1.2.0).
As a major extension, 
it allows now the predictions of (singly) charged
Higgs sectors to be compared against 
current experimental limits from the LEP and Tevatron experiments. 
We have also
significantly extended the variety of neutral Higgs searches included. 
From the LEP experiments, we newly included searches where the Higgs boson
is assumed to decay invisibly or into $\gamma\gamma$ or (non flavour
tagged) hadrons.
Also newly included are the decay-mode independent search of
OPAL \cite{hep-ex/0206022} and several searches involving the production
process $e^+ e^- \to f\bar f H$ with $f\in\{b, \tau\}$.
From the Tevatron experiments, we newly included the analyses released
publicly at the time of the ICHEP 2010 Conference, thus replacing many
implemented analyses by their updated versions. 
Furthermore, we newly 
include the CDF and D0 combination 
of their searches for the individual signal topology
exclusive neutral Higgs boson production
with decay into  $W^+W^-$ \cite{arXiv:1005.3216}, 
which is applicable to a wide class of 
models beyond the SM, a search via the $t\bar tH$ production process 
of \cite{D05739}
and a search assuming the decay $H \to \gamma Z$ \cite{arXiv:0806.0611}.
A full list of the experimental analyses 
and the respective references
included in {\tt
HiggsBounds} \vers{2}{0}{0} is given in \refse{sec:listofanalyses}.

In this release, we do not yet include charged Higgs search channels
which involve the decay of charged Higgs bosons to neutral ones
plus a $W$ boson $H_i^\pm \to h^0_j W^\pm$.
Neither are search results for doubly charged Higgs bosons included.
Therefore, we will refer to the singly charged Higgs bosons
below just as charged Higgs bosons.

The rest of the paper is organised as follows.
In Section~\ref{sec:general approach}, we give a brief description 
of the general approach of the program \HB \vers{2}{0}{0}.
Section \ref{sec:inputdescription} describes the
model predictions required by 
the program as input
in order to enable the application of Higgs search results.
The three input options of \HB\ are described in detail, illustrating
the input required from the user in each case and indicating how this
input is further processed.
Section~\ref{sec:listofanalyses} lists all analyses and their
corresponding references implemented in version 2.0.0 of the 
program. Furthermore, the conditions implemented in the code
which are responsible for determining the applicability of some analyses
to certain Higgs bosons are described here.
The complete operating instructions of \HB \vers{2}{0}{0} 
for the library of subroutines, the command-line and the 
online version with examples are given in Section \ref{sec:usage}.
In Section~\ref{sec:applications}, we present two new applications 
of the program. One application considers a SM-like Higgs boson where 
an extra Higgs decay channel to invisible particles is open;
the other studies the scalar sector of the Randall-Sundrum
model and shows, to our knowledge, for the first time the impact
of Tevatron Higgs search results on this model.

\section{General Approach}
\label{sec:general approach}

The following is a short, but self contained, description of the 
basic work-flow of {\tt HiggsBounds}. A more detailed treatment,
e.g. of the limitations of the methods employed here, can be
found in the manual of 
\HB\ \vers{1}{0}{0}\ \cite{arXiv:0811.4169}.

The user provides the Higgs sector predictions of the model under
consideration. For each 
neutral or charged
Higgs boson $H_i \;  (i=1,\ldots, \nHneut + \nHplus)$  in the model,
this 
includes
information on the Higgs mass, Higgs total decay width, Higgs and top quark
branching ratios and Higgs production cross sections,
\BEA
\label{basic input}
& M_{H_i} \,, 
\Gamma_{\TOT}(H_i)\,,  
\BR_\MOD(H_i\to ...)\,, \BR_\MOD(t\to ...)\,,
 \frac{\sigma_\MOD(P)}{\sigma_{\REF}(P)}&\, ,
\EEA%
and information whether each neutral Higgs boson is CP-odd, CP-even or 
is not a CP eigenstate.
$P$ 
stands for
the Higgs production process. A full list 
of the required Higgs sector predictions
is given in
\refse{sec:inputdescriptionhadr}. Where it exists, $\sigma^\SM(P)$ is used
as the reference production cross section.  Variations on 
the input format (\ref{basic input})
are offered, as described in detail in \refse{sec:inputdescription}, which
allow the user to specify a simpler set of input, 
provided
certain basic approximations are valid.

Note that the user may, to some extent, use the program also with 
incomplete model information by setting not provided cross section ratios,
branching ratios or effective couplings to zero. 
However, the program may then consider some analyses not 
applicable, which would be taken into account if more complete 
model information had been provided.

The {\tt HiggsBounds} package includes sample files which demonstrate how
the code
can be used in conjunction with the widely used 
programs for predictions in the MSSM Higgs sector
\feynhiggs\ and \cpsuperh.
The possibility to link \HB\ with the codes 
\darksusy, \fittino, \mastercode, \THDMC\ has already been demonstrated
in various applications.

A list of the experimental analyses currently included in
{\tt HiggsBounds} \vers{2}{0}{0} can be generated by running the code.
These include results from both LEP and the Tevatron and consist of tables
of expected cross section limits
(based on Monte Carlo simulations with no signal) and observed ones
at 95\% C.L., with a variety of normalisations. The list
mainly consists of analyses for which model-independent limits were
published. However, we also include some dedicated analyses carried out for
the case of the SM, or for Higgs bosons with certain CP properties. These
analyses can only be applied to Higgs bosons which show the
corresponding characteristics to a sufficiently high level of accuracy.

We call the application of a Higgs search analysis to a
particular Higgs boson 
(or two Higgs bosons, if the analysis involves two of them)
of the model under 
study with particular mass(es) an ``analysis application'' $X$.
\footnote{As an example,
suppose a model with three neutral Higgs bosons ($h_1$, $h_2$, $h_3$) 
should be checked against two neutral Higgs search analyses 
$A_1$ and $A_2$, then there are
six possible analysis applications $X$ ($A_1$ applied 
to $h_1$, $h_2$, $h_3$ and $A_2$ applied to $h_1$, $h_2$, $h_3$)
that the model can be tested against.}
To each analysis application $X$ corresponds a signal cross section 
prediction $\sigma(X)$ for the particular Higgs boson
on which an upper limit is put by the analysis.
For each $X$, {\tt HiggsBounds} uses the input to calculate
the relevant quantity $Q_\MOD(X)$ in terms of which the limit is given
(i.e., a conveniently normalised cross section $\sigma(X)$ 
times branching ratio).

The normalisation is carried out using SM predictions for 
Higgs boson production cross sections
and decay branching ratios from
{\tt HDECAY}~\citehdecay version 3.4, 
the TEV4LHC Higgs Working Group~\cite{hep-ph/0612172} 
(for a full list of references, see Tab. \ref{table:SMfunctions}), 
{\tt VFB@NLO}~\cite{hep-ph/0306109},
{\tt HJET}~\cite{hep-ph/0305321,arXiv:0705.2744,arXiv:1003.4438} 
version 1.3 and
dedicated calculations of our own~\cite{arXiv:0811.4169}. 

The rationale behind the choice of SM normalisation 
is that virtually all Tevatron analyses
implemented in this program use these predictions when normalising 
their cross section limits to SM quantities. 
Thus, describing deviations from the SM of a new model by 
using the SM normalisation the experimental analyses have chosen 
allows for the most accurate interpretation of the limits.
By using SM predictions 
for Higgs production cross sections which deviate 
from the ones used in the experimental analyses by a certain amount
(be it because of a different loop order, 
different numerical values of input parameters,
renormalisation scheme or choice
of parton distribution functions),
a deviation of the same relative size will be caused in 
the quantities $Q_\MOD$.
The user should bear this in mind when interpreting the output of 
{\tt HiggsBounds}.

In order to ensure the correct statistical interpretation of the results, it
is crucial to only consider the experimentally observed limit for one
particular analysis application. 
Therefore, {\tt HiggsBounds} must first determine $X_0$,
which is defined as the 
analysis application
with the highest statistical sensitivity
for the model point under consideration. In order to do this, the program
uses the tables of expected experimental limits to obtain a quantity
$Q_\EXPEC$ corresponding to each $X$. 
The 
analysis application
with the largest value of
$Q_\MOD/Q_\EXPEC$ is chosen as $X_0$.

{\tt HiggsBounds} then 
determines
a value for $Q_\OBS$ for this process $X_0$,
using the appropriate table of experimentally observed limits. 
If
\begin{equation}
\frac{Q_\MOD(X_0)}{Q_\OBS(X_0)} > 1
\,,
\label{eq:modvsobs}
\end{equation} 
{\tt HiggsBounds} returns as a result
that this particular parameter point is excluded
at 95 \% C.L. \footnote{ 
Note that if we had compared predicted cross sections
directly against the experimentally measured limits for {\em all} available search
channels and had excluded the model if at least one of them excludes
it at 95~\% C.L., the derived constraint would in general not correspond to a
constraint at 95~\% C.L.
The maximally 5~\% probability of each
individual comparison of $Q_\MOD$ and $Q_\OBS$ to yield a false exclusion
would yield an overall probability for false exclusion which is higher
than 5~\%.}

In order to use {\tt HiggsBounds}, the narrow-width approximation must be
valid for each Higgs boson described in the input,
because the
experimental exclusion bounds currently utilised 
in the program
have all been obtained under this 
assumption.
HiggsBounds can be used
with models which do not change the signature of the background processes
considerably\footnote{ However, new physics models which show strong
deviations from the SM in the background processes to Higgs production are
not usually considered in the literature, since this would often put them in
conflict with SM electroweak precision
data~\cite{hep-ex/0509008,hep-ex/0612034}.}. In addition, models should not
significantly change the kinematical distributions of the signal 
cross section associated with
$X$ (e.g.\ $\eta$, $p_T$ distributions of the final state particles) from
that assumed in the corresponding analysis (for a full discussion see \cite{arXiv:0811.4169}). 

In addition, the LEP Higgs vector boson fusion (VBF) cross section should not be significantly 
enhanced compared to the LEP Higgsstrahlung cross section. (Models with a common 
$g_{h_iVV}/g^{\SM}_{HVV}$ coupling for $V=Z,W$ and no
new mediating particles, such as the MSSM, satisfy this condition automatically.)

\section{Theoretical predictions required as input}
\label{sec:inputdescription}

The user can choose between three different input formats, labelled by the
variable {\tt whichinput}.  
We describe here in detail what is required from the user for each of 
the three settings ({\tt hadr}, {\tt part} and {\tt effC}).
A summary can be found in \refta{table:instructions2}, 
Sect.~\ref{subsec:Common features: Input}.

\subsection{The most general input form: {\tt whichinput=hadr}}
\label{sec:inputdescriptionhadr}

The option {\tt whichinput=hadr} requires the model input in the most general form. It involves specifying (at most)
\begin{enumerate}
\item masses for the neutral Higgs bosons $h_k \; (k=1,\nHneut)$ and singly charged Higgs bosons $H_j^{\pm} \; (j=1,\nHplus)$,
\BEA
m_{h_k}\,,m_{H_j^{\pm}},\non
\EEA
\item Higgs total decay widths,
\BEA
\Gamma_\TOT(h_k)\,,\Gamma_\TOT(H_k^{\pm}),\non
\EEA
\item whether the neutral Higgs boson is CP-even, CP-odd or mixed CP,
\item neutral Higgs branching ratios with SM equivalents (OP=ordinary
	particles i.e. particles which exist in the SM),
\BEA
\BR_\MOD(h_k\to \text{OP}) \:{\rm with}\:
\text{OP = $s\bar{s}$, $c\bar{c}$, $b\bar{b}$, $\tau^+\tau^-$,
 $W^+W^-$, $ZZ$, $Z\gamma$, $\gamma\gamma$, $gg$,}\non
\EEA
\item neutral Higgs branching ratios without SM equivalents
\BEA
\BR_\MOD(h_k\to h_i h_i),\BR_\MOD(h_k\to {\rm invisible}),\non
\EEA
\item charged Higgs branching ratios to SM particles 
\BEA
\BR(H_j^{+} \to {\text{OP}})\:{\rm with}\:
\text{OP = $c\bar{s}$, $c\bar{b}$, $\tau^+ \nu_{\tau}$,}\non
\EEA
\item top quark branching ratios
\BEA
\BR(t \to W^{+}b),\BR(t \to H_j^{+}b),\non
\EEA
\item normalised cross section $R_{\sigma}(P)$ for LEP Higgs production process $P$
\BEA 
 & &  e^+e^- \to h_j Z              ,\non\\
 & &  e^+e^- \to b \bar{b} h_j      ,\non\\
 & &  e^+e^- \to \tau^+ \tau^- h_j  ,\non\\
 & &  e^+e^- \to h_j h_i            ,\non\\
 & &  e^+e^- \to H^+_j H^-_j        ,\non
\EEA
\item normalised cross section $R_{\sigma}(P)$ for Tevatron hadronic Higgs
	production process $P$
\BEA 
 & &  p\bar{p} \to h_j   ,\non\\
 & &  p\bar{p} \to b h_j ,\non\\
 & &  p\bar{p} \to h_j W ,\non\\
 & &  p\bar{p} \to h_j Z ,\non\\
 & &  p\bar{p} \to h_j  {\rm \, via \, VBF}  ,\non\\
 & &  p\bar{p} \to t \bar{t} h_j             .\non
\EEA
\end{enumerate}

Note that only a subset of these inputs may be required. 
For example, if 
only
the LEP limits are required by the user, no Tevatron cross
sections will need to be given as input (see \refse{sec:usage}). If the user
only wishes to test the neutral Higgs sector, no input involving the charged
Higgs sector or top decays will be required.

The normalised cross section of a Higgs production process $P$ is defined by
\BEA 
\label{eq:R-sigma}
R_{\sigma}(P)&=&\frac{\sigma_{\rm model} (P)}{\sigma_{\rm ref} (P)} \non
\EEA
Where the SM equivalent exists, the reference cross section
$\sigma_{\rm ref} (P)$ of a process $P$ involving a Higgs boson $h_k$ is
$\sigma_{\rm ref} (P)=\left.\sigma_{\rm SM} (P)\right|_{m^{\rm SM}_H=m_{h_k}}$.

The only neutral Higgs production process without a SM
equivalent considered in \HB\ up to now
is $P=e^+e^- \to h_j h_i$. In this case, as in \HB\ \vers{1}{0}{0}\ 
\cite{hep-ex/0602042}, we choose a reference cross section for a fictitious
production process of two scalar particles ($H'$, $H$) with masses
$m_{H'}=m_{h_j}$ and 
$m_H = m_{h_i}$ via a virtual
$Z$ exchange with a standardised squared coupling constant: 
\begin{align} 
\label{eq:gref-HHZ}
\left(g^{\REF}_{H'HZ}\right)^2 =\frac{e^2}{4 \sw^2 \cw^2}
\,,
\end{align}
where $e$ denotes
the electromagnetic coupling constant, and $\sw$ and $\cw$
the sine and cosine
of the weak mixing angle, respectively.

The cross section of this process in leading order is then completely 
determined by the Higgs masses and SM input and related
to the SM Higgsstrahlung cross section via a simple phase space factor:
\begin{align}
\label{eq:eeHH-ref}
\sigma_\REF(\text{$H'\, H$ production})&
	=\bar{\lambda}\left(m_{H'},m_{H},s\right)
         \sigma^{\rm SM}_{HZ}(m_H) \,,\\
\nonumber
\bar{\lambda}\left(m_{H'},m_{H},s\right)&
	=\frac{\lambda^{3/2}_{H'H}(s)}
        {\lambda^{1/2}_{HZ}(s)\left(\lambda_{HZ}(s)+12
         \frac{m_Z^2}{s}\right)} \,,\\
\nonumber
\lambda_{ab}(s)&=\left[1-\frac{\left(m_a+m_b\right)^2}{s}\right]
                 \left[1-\frac{\left(m_a-m_b\right)^2}{s}\right] \,.
\end{align}
This reference cross section coincides with the MSSM tree level cross
section for the process $e^+ e^- \to h^0 A^0$ if the Higgs mixing-angle
dependent factor $\cos(\beta-\alpha)$ is divided out of the tree level
coupling and $m_{A^0}$ and $m_{h^0}$ are chosen as $m_{H'}$ and $m_{H}$.
Therefore, $R_{\sigma}(e^+ e^- \to h^0 A^0)$ is simply given by
$\cos^2(\beta-\alpha)$ in the MSSM (with real parameters) at tree level.

For the process $P=e^+e^- \to H^+_j H^-_j$, the reference cross section is
the cross section of the process $e^+e^- \to H^+ H^-$ 
in a 2 Higgs doublet model (e.g. the MSSM) at tree-level 
(i.e. s-channel $\gamma$ and $Z$ exchange). 
This reference cross section depends solely on
the mass of the charged Higgs boson and SM quantities.
As a consequence,
in the MSSM, $R_{\sigma}(e^+e^- \to H^+_j H^-_j)=1$.

The invisible branching ratio $\BR_\MOD(h_k\to {\rm invisible})$ is defined
as the branching ratio of a neutral Higgs boson into particles which only
appear in the detector as missing energy. Examples of invisible particles
include neutralinos in the MSSM or majorons in supersymmetric 
models with spontaneous breaking of R-parity.

The top quark branching ratios $\BR(t \to H_j^{+}b)$ and $\BR(t \to W^{+}b)$
are also required as input, when the model is compared with the charged
Higgs searches from the Tevatron. $\BR(t \to W^{+}b)$ is required for
analyses in order to determine when analyses which assume $\BR(t \to
H_j^{+}b)+\BR(t \to W^{+}b)\sim 1$ can be applied to the model point.

In order to make it more convenient for users to calculate this input, the
{\tt HiggsBounds} library provides a series of Fortran functions which allow
the user to access the predictions of 
certain SM quantities, including the SM Tevatron hadronic
Higgs production cross sections, total decay width and branching ratios as a
function of Higgs mass and $\Gamma^{\rm SM}(t \to W^{+}b)$ as function of the
top mass. These are listed in \refta{table:SMfunctions}
together with the corresponding references.

\subsection{Specifying some cross sections at the partonic level: {\tt whichinput=part}}

The input required for the option {\tt whichinput=part} 
may be
more convenient for the user to calculate than for the option 
{\tt whichinput=hadr}, 
since 
most
of the Tevatron cross section ratios can be
specified at the partonic level rather than the hadronic level.

The input option {\tt whichinput=part} requires, at most,
\begin{enumerate}
\item[(1)-(8)] as for {\tt whichinput=hadr}
\item[(9)] normalised cross section $R_{\sigma}(P)$ for Tevatron hadronic neutral Higgs production process $P$
\BEA 
 & &  p\bar{p} \to h_j  {\rm \, via \, VBF}\non\\
 & &  p\bar{p} \to t \bar{t} h_j \non
\EEA
\item[(10)] normalised cross section $R^{h_j+y}_{nm}$ for Tevatron partonic 
 neutral Higgs production process $nm\to h_j+y$,
 where $y$ indicates a particle produced in association with the 
 Higgs boson (or none if $y$ is omitted):
\BEA
 R^{h_j}_{nm},                &\,{\rm with}\, nm = & gg ,b\bar{b} ,      \non\\
 R^{h_j +W^+}_{nm},           &\,{\rm with}\, nm = & u\bar{d}, c\bar{s}, \non\\
 R^{h_j +W^-}_{nm},           &\,{\rm with}\, nm = & d\bar{u}, s\bar{c}, \non\\
 R^{h_j +Z}_{nm},             &\,{\rm with}\, nm = & d\bar{d},u\bar{u},s\bar{s},c\bar{c},b\bar{b}, \non\\
 R^{h_j +b,h_j +\bar{b}}_{nm},&\,{\rm with}\, nm = & bg,\bar{b}g \non
\EEA
\end{enumerate}

The normalised cross section $R^{h_j+y}_{nm}$ for a Tevatron partonic
neutral Higgs production process
is defined by
\BEA
\label{eq:R-nm-partonic}
R^{H+y}_{nm}&=&\frac{\hat{\sigma}^{\rm model}_{nm\to H+y} }{\hat{\sigma}^{\rm SM}_{nm\to H+y} }
\EEA
calculated for a parton-system centre-of-mass energy squared $\hat{s}=\hat{s}_0$, where $\hat{s}_0$ denotes the partonic production threshold $\hat{s}_0=\left(m_H+m_y\right)^2$ (with $m_y=0$ in the case of single Higgs boson production). This requires that the dependence on $\hat{s}$ is (at most) mild.

Internally, {\tt HiggsBounds} uses the relation
 \begin{align}
\label{eq:R-sigma-approx}
R_\sigma(P) & \approx \sum_{\{n,m\}}
	R_{nm}^{H+y}
	\frac{
	  \sigma_\SM(p\bar p \to n m \to H+y)
	}{\sigma_\SM(p\bar p \to H+y)}\,,
\end{align}
to calculate the hadronic cross section ratios from the partonic cross
section ratios. The ratios 
$\frac{\sigma_\SM(p\bar p \to n m \to H+y)}{\sigma_\SM(p\bar p \to H+y)}$ 
are provided within {\tt HiggsBounds}.
(For further explanation of the applicability of this approximation and
details of how $\frac{\sigma_\SM(p\bar p \to n m \to H+y)}{\sigma_\SM(p\bar p \to H+y)}$ 
are calculated, see \cite{arXiv:0811.4169}).

The partonic cross section ratios $R^{H+y}_{nm}$ are much easier to calculate than hadronic cross section ratios in a wide range of models. In addition, it is often possible to make the approximation
\BEA
R^{h_j +W^+}_{nm}=R^{h_j +W^-}_{nm}=R^{h_j +Z}_{nm} =: R^{h_j +V}
\label{eq:gaugebosonpartratios}
\EEA
for all $nm$, thus substantially reducing the number of partonic cross
section ratios which need to be 
provided by the user
from twelve to four.

For instance in the MSSM with real parameters,
this ratio $R^{h_j +V}$ can be calculated approximately 
from the normalised squared effective Higgs coupling to two Z bosons, i.e.
$(g^\MOD_{h_jZZ}/g^\SM_{HZZ})^2$
(for definition of this
coupling, see \refse{section:effC}). Similarly, $R^{h_j}_{b\bar{b}}$ and
$R^{h_j +b,h_j +\bar{b}}_{bg,\bar{b}g}$ can be taken to be approximately
equal in the MSSM (and can also be calculated approximately from effective
couplings: see \refeq{eq:effCpartc}).

\subsection{The effective coupling approximation: {\tt whichinput=effC}}
\label{section:effC}

The option {\tt whichinput=effC} allows the user to specify a greatly
reduced number of input parameters. It involves specifying (at most)
\begin{enumerate}
\item masses for the neutral Higgs bosons $h_k \; (k=1,\nHneut)$ and 
	singly charged Higgs bosons $H_j^{\pm} \; (j=1,\nHplus)$,
\BEA
m_{h_k}\,,m_{H_j^{\pm}},\non
\EEA
\item Higgs total decay widths,
\BEA
\Gamma_\TOT(h_k)\,,\Gamma_\TOT(H_k^{\pm}),\non
\EEA
\item normalised squared scalar and pseudoscalar effective couplings to fermions 
\BEA
\left(\frac{g^\MOD_{s,h_k(\text{OP})}}{g^\SM_{H(\text{OP})}}\right)^2\,,
\left(\frac{g^\MOD_{p,h_k(\text{OP})}}{g^\SM_{H(\text{OP})}}\right)^2\,,
\text{ OP = $s\bar{s}$, $c\bar{c}$, $b\bar{b}$, $t\bar{t}$, $\tau^+\tau^-$}
\non
\EEA
\item normalised squared effective couplings to bosons 
\BEA
\left(\frac{g^\MOD_{h_ih_jZ}}{g^{\REF}_{H'HZ}}\right)^2\,,
\left(\frac{g^\MOD_{h_k(\text{OP})}}{g^\SM_{H(\text{OP})}}\right)^2\,,
\text{ OP = $W^+W^-$, $ZZ$, $Z\gamma$, $\gamma\gamma$, $gg$}\non
\EEA
\item neutral Higgs branching ratios without SM equivalents, charged Higgs
	branching ratios to SM particles and top quark branching ratios as before
\end{enumerate}
From this input, the input corresponding to the option {\tt part}
is calculated.

We define the scalar and pseudoscalar Higgs coupling to fermions in the
usual way, via the Feynman rule for the coupling of a generic neutral Higgs
boson $h$ to fermions:
\begin{align}
G(h f \bar f) & = i(g_s \text{\bf 1} + i g_p \gamma_5)
\,,
\end{align}
where $g_s$ and $g_p$ are real-valued scalar and pseudoscalar coupling
constants respectively, and $\text{\bf 1}$ and $\gamma_5$ are the usual
matrices in Dirac space.
A scalar particle, like the SM Higgs boson, has $g_p=0$
and a pseudoscalar particle has $g_s=0$. 

Where it exists, the reference couplings squared $(g^\SM_{H(\text{OP})})^2$ are the
SM tree-level equivalents:
\begin{align}
\label{eq:gref-HZZ}
(g^\SM_{HZZ})^2 & = 
	\left(\frac{e}{\sw}\frac{m_W}{\cw^2}\right)^2\,,\\
\label{eq:gref-HWW}
(g^\SM_{HWW})^2 & = 
	\left(\frac{e}{\sw}\: m_W\right)^2\,,\\
\label{eq:gref-Hff}
(g^\SM_{Hf\bar f})^2 & = 
	\left(\frac{1}{2}\frac{e}{\sw}\frac{m_f}{m_W}\right)^2\,,
\end{align}
where $m_W$ and $m_f$ denote the masses of the $W$~boson 
and fermion $f$, respectively. $(g^{\REF}_{H'HZ})^2$ 
is defined in \refeq{eq:gref-HHZ} above.

The effective couplings 
$(g^\MOD_{h_k (\text{OP})}/g^\SM_{H (\text{OP})})^2$,
with OP = $Z\gamma$, $\gamma\gamma$, 
are loop-induced. They can be defined via
\BEA
\left(\frac{g^\MOD_{h_k (\text{OP})}}{g^\SM_{H (\text{OP})}}\right)^2
	&=& \frac{\Gamma^\MOD_{h_k\to \text{OP}}(m_{h_k})}
		{\Gamma^\SM_{H\to \text{OP}}(m_H)|_{m_H=m_{h_k}}}
\EEA
There is a choice of definition of the Higgs-gluon-gluon effective coupling
squared 
$(g^\MOD_{h_k (\text{OP})}/g^\SM_{Hgg})^2$.
It can either be defined via decay widths:
\BEA
\left(\frac{g^\MOD_{h_k (\text{OP})}}{g^\SM_{Hgg}}\right)^2
	&=&\frac{\Gamma^\MOD_{h_kgg}(m_{h_k})}
		{\Gamma^\SM_{H\to \text{OP}}(m_H)|_{m_H=m_{h_k}}}
\label{eq:defggheffCa}
\EEA
or via partonic cross sections:
\BEA
\left(\frac{g^\MOD_{h_k (\text{OP})}}{g^\SM_{Hgg}}\right)^2&=&R_{g g}^{h_k}.
\label{eq:defggheffCb}
\EEA
In general, the input option {\tt whichinput=effC} should only be used when
both definitions result in very similar values for 
$(g^\MOD_{h_k(\text{OP})}/g^\SM_{Hgg})^2$. 
However, in certain circumstances, this
condition can be relaxed. For example, in a model in which the 
LEP searches for Higgs bosons decaying into hadrons
are not relevant, the normalised gluon-gluon-Higgs effective
coupling can be defined solely by \refeq{eq:defggheffCb}. Conversely, if the
Tevatron gluon fusion Higgs production mechanism is not relevant, the
normalised gluon-gluon-Higgs effective coupling can be defined solely by
\refeq{eq:defggheffCa}.

The normalised LEP cross sections are calculated from the normalised
effective couplings using the relations:
\begin{align}
\label{eq:effCLEPa}
R_{\sigma}(e^+e^-\to h_k Z)& =
\left(\frac{g^\MOD_{h_kZZ}}{g^\SM_{HZZ}}\right)^2
\,, \\
\label{eq:effCLEPb}
R_{\sigma}(e^+e^-\to h_k h_i)& =
	\left(\frac{g^\MOD_{H'H Z}}{g^\REF_{H'HZ}}\right)^2
\,, \\
\label{eq:effCLEPc}
R_{\sigma}(e^+e^-\to b \bar{b}h_k^{\rm CP\,even} )& =
\left(\frac{g^\MOD_{s,h_k b\bar b}}{g^\SM_{H b\bar b}}\right)^2
\,, \\
\label{eq:effCLEPd}
R_{\sigma}(e^+e^-\to b \bar{b}h_k^{\rm CP\,odd} ) & =
\left(\frac{g^\MOD_{p,h_k b\bar b}}{g^\SM_{H b\bar b}}\right)^2
\,, \\
\label{eq:effCLEPe}
R_{\sigma}(e^+e^-\to \tau^+\tau^- h_k^{\rm CP\,even} ) & =
\left(\frac{g^\MOD_{s,h_k \tau^+\tau^-}}{g^\SM_{H \tau^+\tau^-}}\right)^2
\,, \\
\label{eq:effCLEPf}
R_{\sigma}(e^+e^-\to \tau^+\tau^- h_k^{\rm CP\,odd} )& =
\left(\frac{g^\MOD_{p,h_k \tau^+\tau^-}}{g^\SM_{H \tau^+\tau^-}}\right)^2\,.
\end{align}
The analyses currently in {\tt HiggsBounds} which use $\sigma_\MOD(e^+e^-\to b \bar{b}h_k )$ or $\sigma_\MOD(e^+e^-\to \tau^+\tau^- h_k )$ apply only to 
Higgs bosons which are CP eigenstates.

The partonic Tevatron cross section ratios are calculated by 
\begin{align}
\label{eq:effCparta}
R_{g g}^{h_k}&=
	\left(\frac{g^\MOD_{h_kgg}}{g^\SM_{Hgg}}\right)^2
\,,\\
\label{eq:effCpartb}
R_{b \bar b }^{ h_k}
	=
	R_{b g,\bar{b} g}^{h_kb,h_k \bar{b}} 
	&=\left(\frac{g^\MOD_{s,h_k b\bar b}}{g^\SM_{H b\bar b}}\right)^2+
\left(\frac{g^\MOD_{p,h_k b\bar b}}{g^\SM_{H b\bar b}}\right)^2
\,, \\
\label{eq:effCpartc}
R_{q \bar q' }^{ h_k W^+}
	= 
R_{q' \bar q }^{ h_k W^-}
	&=
	\left(\frac{g^\MOD_{h_kWW}}{g^\SM_{HWW}}\right)^2
\,,\\
\label{eq:effCpartd}
R_{q'' \bar q'' }^{ h_k Z}
	&=
	\left(\frac{g^\MOD_{h_kZZ}}{g^\SM_{HZZ}}\right)^2
\,,
\end{align}
where $(q,q')\in\{(u,d),(c,s)\}$ and $q''\in\{u,d,c,s,b\}$.

The $t \bar t$ CP-even Higgs hadronic Tevatron cross section ratio is obtained using 
\begin{align}
\label{eq:effCtth}
R_{\sigma}(p \bar p \to t \bar t  h^{\rm CP\,even}_k )
	&=
	\left(\frac{g^\MOD_{s,h_k t \bar t}}{g^\SM_{H t \bar t}}\right)^2
\,.
\end{align}

The normalised hadronic cross sections for $h_k$ production via VBF is calculated using the approximate relation:
\begin{align}
\label{eq:effCvbf}
R_{\sigma}(p\bar p \to h_k \text{ via VBF})
	&=
 	 R_{\text{VBF}}^{WW}\,
		\left(\frac{g^\MOD_{h_kWW}}{g^\SM_{HWW}}\right)^2
	+R_{\text{VBF}}^{ZZ}\,
		\left(\frac{g^\MOD_{h_kZZ}}{g^\SM_{HZZ}}\right)^2
\,.
\end{align}
where the numbers
\begin{align}
R_{\text{VBF}}^{WW} & :=\frac{\sigma_\SM(p\bar p \to H \text{ via $WW$ fusion})}
        {\sigma_\SM(p\bar p \to H \text{ via VBF})}
	= 77\% 
\,,\\
R_{\text{VBF}}^{ZZ} &:= \frac{\sigma_\SM(p\bar p \to H \text{ via $ZZ$ fusion})}
        {\sigma_\SM(p\bar p \to H \text{ via VBF})}
	= 23\%
\,.
\end{align}
have been calculated for $p\bar p$ collisions with 1.96 TeV centre-of-mass
energy using VBFNLO \cite{hep-ph/0306109}. Including the Higgs mass
dependence in the range $70\,\gev < m_H < 300\,\gev $ would change these
proportions by less than 1\%. Interference effects
also affect the result by less than 1\% (cf. \cite{hep-ph/0503172}). (See \citere{arXiv:0811.4169} for a more detailed discussion.).

Within {\tt HiggsBounds}, the neutral Higgs decay widths to ordinary
particles in the effective coupling input option are then calculated by
\BEA
\label{eq:effCgammaa}
\Gamma^\MOD_{h_k\to \text{OP}}(m_{h_k})
&=&
\left(\frac{g^\MOD_{h_k (\text{OP})}}{g^\SM_{H (\text{OP})}}\right)^2\left.\Gamma^\SM_{H\to \text{OP}}(m_H)
\right|_{m_H=m_{h_k}}\non\\
&&\text{for OP = $W^+W^-$, $ZZ$, $Z\gamma$, $\gamma\gamma$, $gg$,}\\
\label{eq:effCgammab}
\Gamma^\MOD_{h_k\to \text{OP}}(m_{h_k})
&=&
\left(\left(\frac{g^\MOD_{s, h_k (\text{OP})}}{g^\SM_{H (\text{OP})}}\right)^2
	+\left(\frac{g^\MOD_{p, h_k (\text{OP})}}{g^\SM_{H (\text{OP})}}\right)^2
\beta_f^{-2}(m_{h_k})
\right)
\left.\Gamma^\SM_{H\to \text{OP}}(m_H)
\right|_{m_H=m_{h_k}}\non\\
&&\text{for OP = $s\bar{s}$, $c\bar{c}$, $b\bar{b}$, $\tau^+\tau^-$,}
\EEA
with $\beta_f^2(m_{h_k}) = 1-4m_f^2/m_{h_k}^2$.
These are converted to branching ratios via
\BEA
\label{eq:effCbr}
\BR^\MOD_{h_k\to \text{OP}}(m_{h_k})
	&=&\frac{\Gamma^\MOD_{h_k\to \text{OP}}(m_{h_k})}{\Gamma_\TOT(h_k)}
\EEA
Note that this means that it is especially important for the user to give an
accurate $\Gamma_\TOT(h_k)$ when using this input option.

Using effective couplings as input, the CP properties of the neutral 
Higgs bosons can be inferred, at least as far as the signal properties
considered here are concerned.
In the present implementation, we set the CP value of a Higgs $h_k$ in the 
following way:
\begin{align*}
\text{CP}(h_k) \to +1 & && \text{if}\;\; \max_{f}
\left(\frac{g^\MOD_{p,h_k f\bar f}}{g^\SM_{H f\bar f}}\right)^2 < 10^{-16},\\
\text{CP}(h_k) \to -1 & && \text{if}\;\; \max_{f}
\left(\frac{g^\MOD_{s,h_k f\bar f}}{g^\SM_{H f\bar f}}\right)^2  < 10^{-16},\\
\text{CP}(h_k) \to 0 & && \text{if neither of the above conditions is met.}
\end{align*}

For the implemented set of analyses, the limits
on topological cross sections are applicable with high accuracy
to models where the Higgs sector can be faithfully parametrised with 
effective couplings (See \citere{arXiv:0811.4169} for 
details).

In summary, users may find it convenient to use the input option 
{\tt whichinput=effC} for cases in which \refeq{eq:R-sigma-approx}, 
Eqs.~(\ref{eq:effCLEPa}) --
(\ref{eq:effCvbf}),
and 
Eqs. (\ref{eq:effCgammaa}) and (\ref{eq:effCgammab})
are valid to a satisfactory level of accuracy. 
Of course, if not all of the analyses are relevant to the model, 
some of these relations do not need to hold. 
We have already discussed that, in the case of a model in which the 
LEP searches for Higgs bosons decaying into hadrons
are not relevant, \refeq{eq:defggheffCa} does not need
to be valid.

\section{Analyses in {\tt HiggsBounds} \vers{2}{0}{0}}
\label{sec:listofanalyses}
{\tt HiggsBounds} contains many of the available observed and expected
exclusion limits at 95\% C.L. (preliminary and final results) from LEP and
the Tevatron as data tables which are read in by the program during
start-up. In order to provide values for $Q_\OBS(X)$ and $Q_\EXPEC(X)$ for
continuous Higgs mass values, the program interpolates $Q$-values linearly
between neighbouring Higgs mass points. We aim to include not only the most
recent analysis in each channel by each collaboration (or analysis combining
results from more than one collaboration) but also the most recent
result appearing in a paper submitted to the arXiv. For simplicity we
use the label `published' for those results (since results submitted to
the arXiv are usually being prepared for publication in a journal or
conference proceedings).

In \HB\ \vers{2}{0}{0}, in total 81 Higgs search analyses have been implemented
consisting of 28 from LEP and 53 from the Tevatron. 
While many types of analyses have been added for the first time,
several Tevatron analyses included in \HB\ {\tt 1}.{\tt 2}.{\tt 0} have been 
replaced by updated ones based on more data.
For completeness and as a reference for users, 
a full list of the implemented analyses follows.
\newpage 

\noindent {$\bullet$ LEP neutral Higgs analyses 
(considering the final states):}\\
\begin{tabular*}{0.85\textwidth}{@{\extracolsep{\fill}}ll}
$h_k Z, h_k \to b b$ or $h_k \to \tau \tau$ 
	\cite{hep-ex/0602042},
&	  
$h_k Z, h_k \to h_i h_i, h_i \to b b$ \cite{hep-ex/0602042},\\
$h_k Z, h_k \to$ anything 
	\cite{hep-ex/0206022},
&	  
$h_k Z, h_k \to h_i h_i, h_i \to \tau \tau$ \cite{hep-ex/0602042},\\
$h_k Z, h_k \to$ invisible 
	\cite{hep-ex/0107032,hep-ex/0401022,hep-ex/0501033,arXiv:0707.0373},
&	
$h_k h_i,  h_{k,i} \to b b$ \cite{hep-ex/0602042},\\
$h_k Z, h_k \to \gamma \gamma$ 
	\cite{LHWGnotes},
&	
$h_k h_i,  h_{k,i} \to \tau \tau$ \cite{hep-ex/0602042},\\
$h_k Z, h_k \to$ hadrons%
\footnotemark,
&	
$h_k h_i, h_k \to h_i h_i, h_i \to b b$ \cite{hep-ex/0602042},\\
$b\bar b h_k \to b\bar b b\bar b$, $h_k$ CP even or odd, 
	\cite{hep-ex/0410017},
&	
$h_k h_i, h_k \to h_i h_i, h_i \to \tau \tau$ \cite{hep-ex/0602042},\\
$b\bar b h_k \to b\bar b \tau\tau$, $h_k$ CP even or odd, 
	\cite{hep-ex/0410017,hep-ex/0111010},
&	
  $h_k h_i, h_k \to b b, h_i \to \tau \tau$ \cite{hep-ex/0602042},\\
$\tau \tau h_k \to \tau\tau\tau\tau$, $h_k$ CP even or odd, 
	\cite{hep-ex/0410017}.
&	
\end{tabular*}
\smallskip 

\footnotetext{Combination based on
	\cite{hep-ex/0510022,hep-ex/0205055,hep-ex/0312042,hep-ex/0408097}}

\noindent {$\bullet$ Tevatron single topology neutral Higgs analyses 
(considering the final states):}\\
\begin{tabular*}{0.85\textwidth}{@{\extracolsep{\fill}}ll}
$Z h_k \to l l b \bar b$ 
	\cite{CDF10235,arXiv:0908.3534,D06089} ,
	& 
$\text{single } h_k \to W W$
	\cite{arXiv:0809.3930,arXiv:1005.3216} ,
	\\
$W h_k \to l \nu b \bar b$
	\cite{D06092,arXiv:0808.1970,CDF10217,arXiv:0906.5613},
	&
$\text{single } h_k \to \tau \tau$ 
	\cite{arXiv:0805.2491,D05740,arXiv:0906.1014,arXiv:1003.3363},
	\\
$W h_k \to 3 W$, 
	\cite{D05873,CDF7307v3},
	& 
$\text{single } h_k \to Z \gamma$, 
	\cite{arXiv:0806.0611},
	 \\
$b h_k \to 3 b \text{ jets}$
	\cite{CDF10105,D05726,arXiv:0805.3556},
	&
$t\bar th_k \to t\bar tb\bar b$
	\cite{D05739},
	\\
$b h_k \to b \tau \tau$
	\cite{arXiv:0912.0968,D05985,D06083}.
	& 
\end{tabular*}
\smallskip

\noindent {$\bullet$ Tevatron SM Higgs combined analyses
	considering the final states (schematic):}\\
\begin{tabular}{ll}
 $V h_k \to b \bar b + E_T^{\text{miss}}$ with $V\in\{W,Z\}$
	\cite{CDF10212,arXiv:0911.3935,D06087,arXiv:0912.5285},
	& \\
 $V h_k \to VVV \to l^\pm l^\pm + X$ with  $l\in\{e,\mu\}$ and  $V\in\{W,Z\}$ 
	\cite{D06091},
	& \\
 $h_k + X \to W W + X$ 
	\cite{D05757,CDF10102,arXiv:1001.4468,D05871,D06082,arXiv:1001.4481,arXiv:1001.4162},
	& \\
 $h_k + X \to \tau \tau + X$ 
	\cite{CDF9248,D05845,arXiv:0903.4800},
	& \\
 $h_k + X \to b b + X$  
	\cite{CDF10010},
	& \\
 $h_k + X \to \gamma \gamma + X$
	\cite{D05858,arXiv:0901.1887,CDF10065},
	& \\
 $h_k + X$ and various Higgs decays
	\cite{arXiv:0712.0598,arXiv:0712.2383,arXiv:0804.3423,arXiv:0808.0534,%
CDF9999,D06008,arXiv:0911.3930,arXiv:1007.4587}.
\end{tabular} 
\smallskip

\noindent {$\bullet$ LEP/Tevatron charged Higgs analyses:} \\
\begin{tabular}{ll}
$e^+ e^- \to H^+ H^- \to$ 4 jets 
	\cite{hep-ex/0107031,hep-ex/0404012},
	& \\
$e^+ e^- \to H^+ H^- \to \tau\nu\tau\nu$ 
	\cite{hep-ex/0404012},
	& \\
$p\bar p\to t\bar t, t \to H^+ b,\; H^+ \to cs$ (\& c.c.)
	\cite{arXiv:0908.1811,arXiv:0907.1269},
	& \\
$p\bar p\to t\bar t, t \to H^+ b,\; H^+ \to \tau \nu$ (\& c.c.)
   	\cite{arXiv:0908.1811}. 
	&
\end{tabular}

Internally, {\tt HiggsBounds} uses a number of 
predictions for SM quantities
for
the Higgs sector\cite{hep-ph/9704448,hep-ph/0612172,hep-ph/0306109,%
hep-ph/0305321,arXiv:0705.2744,arXiv:1003.4438,PRLTA.70.1372,hep-ph/9504378,%
hep-ph/0201206,hep-ph/0302135,hep-ph/0207004,hep-ph/0306211,hep-ph/0508265,%
hep-ph/0508284,hep-ph/0307206,hep-ph/0306234,hep-ph/0406152,hep-ph/0304035,%
hep-ph/9206246,hep-ph/9905386,hep-ph/0403194,%
hep-ph/0107081,hep-ph/0107101,hep-ph/0211438}
to convert between experimental limits with different normalisations.
\medskip

Some of the analyses have been performed under certain model assumptions.
Typical assumptions are, likeness to the SM, 
or models which fulfil $\BR(t\to W^+ b)+\BR(t\to H_i^+ b)=1$, 
or Higgs bosons with certain CP properties. 
In these cases, \HB\ determines
for a given model scenario
which analyses
can be applied to which Higgs boson.

In order to decide whether a 
model scenario
is sufficiently `SM-like' to be
compared with an analysis carried out under SM assumptions, we
use the following test.
Firstly, for each of the $N_{\text{CS}}$ distinct production cross sections
$\sigma_\MOD(P_i(h))$ and the $N_{\text{BR}}$ distinct decay branching
ratios $\BR(h\to F_k)$ which appear in the list of $M$ signal topologies in
that particular analysis, we determine the normalised mean value and the
deviation from the mean:
\begin{align}
\bar s &= \frac{1}{N_{\text{CS}}} \sum_{i=1}^{N_\text{CS}} s_i
\,, & 
\bar b &= \frac{1}{N_{\text{BR}}} \sum_{k=1}^{N_\text{BR}} b_k
\,, \\
\delta s_i & = s_i - \bar s
\,, &
\delta b_k & = b_k - \bar b
\,,\\
\text{with}\;
s_i & = \frac{\sigma_\MOD(P_i(h))}{\sigma_\SM(P_i(H))}
\,, &
b_k & = \frac{\BR_\MOD(h\to F_k)}{\BR_\SM(H\to F_k)}
\,.
\end{align}

The parameter point is considered `SM-like' for this particular analysis if 
the relative deviations 
of all individual production and decay combinations $(i,k)$
from the mean $\bar s\bar b$ 
stay below a preset bound $\epsilon$:
\begin{align}
\max_{i,k} \left|\frac{\delta s_i}{\bar s}
        + \frac{\delta b_k}{\bar b}
        + \frac{\delta s_i \delta b_k}{\bar s\, \bar b} \right| < \epsilon
\,.
\end{align} 
By default,  $\epsilon = 2\%$, i.e.\ 
the predictions for the different topological cross sections, normalised
to the SM values, 
are required to 
result in the same common scale factor
with at least 2\% accuracy. 
This method is conservative and errs on the side of caution, but can
occasionally be overly restrictive. 

After ensuring that a certain
parameter point passes the SM-likeness test for this
analysis, the normalised theoretical cross section $Q_\MOD$ is
calculated simply by
\BEA
Q_\MOD &=& \bar s \bar b.
\EEA

If two or more neutral Higgs bosons have similar masses, then {\tt
HiggsBounds} allows the possibility of adding up their cross sections. If
the mass difference between two neutral Higgs bosons is less than {\tt
delta\_Mh\_LEP} (or {\tt delta\_Mh\_TEV}), then the full LEP (or Tevatron)
cross sections will be added. The normalisation factor is calculated at the
average mass of the Higgs bosons involved. Note that:
\begin{quote}
{\tt delta\_Mh\_LEP} and {\tt
delta\_Mh\_TEV} are set to zero by default.
\end{quote}
To switch this feature on, the
user should change these variables 
in the code and recompile
(see the readme files 
about where to find these variables in the code). 
This should only be done if the user is confident
that interference effects are small. We would recommend that 
{\tt delta\_Mh\_LEP} $\leq 2 \gev$ and {\tt delta\_Mh\_TEV} $\leq 10 \gev$. For analyses
which require a SM-likeness test, the full cross sections are not added,
regardless of the values of {\tt delta\_Mh\_LEP} and {\tt delta\_Mh\_TEV}.

This feature can be very useful in, for example the $m_h^\MAX$ scenario
of the MSSM \cite{hep-ph/0202167}, where a
CP even Higgs boson and the CP odd Higgs boson can have very similar masses
but, naturally, they do not mix. In this case, the cross sections are
commonly added, such as for example in \cite{CDF10105}
and \cite{D05726}.

\section{New {\tt HiggsBounds} Operating Instructions}

Features which are new in version 2.0.0 will be labelled with the symbol: \mystar.
\label{sec:usage}

There are three formats in which the program {\tt HiggsBounds} can be used:

\begin{itemize}
\item {Library of subroutines}
\item {Command-line version} 
\item {Online version}
\end{itemize}

The most widely applicable format of {\tt HiggsBounds} is the
command-line version, since this reads all the model data from text
files and thus this model data can be generated using any package the
user wishes. The library of {\tt HiggsBounds} subroutines
allows {\tt HiggsBounds} to be 
called within other programs.
If
the user just wishes
to check a few parameter points, the online version provides
quick access to all the functionality of {\tt HiggsBounds}, without the
need to 
install 
the code. 

The {\tt HiggsBounds} code, the online version and documentation can all
be found at the URL {\tt http://projects.hepforge.org/higgsbounds/} .
\footnote{The former website {\tt www.ippp.dur.ac.uk/HiggsBounds} should
redirect to the new one.}

The {\tt HiggsBounds} code is provided in either Fortran 77 or Fortran 90. 
Both codes provide exactly the same functionality and have exactly the
same operating instructions. 
In fact, the Fortran 77 and Fortran 90 versions of the {\tt HiggsBounds}
subroutines can even be called within codes written in Fortran 90 
and Fortran 77, respectively.
Both codes have also been tested with a variety of Fortran
compilers, including the free gnu compilers which accompany most Linux
distributions. Therefore, the user may 
install
either code and the
difference will only be apparent if the user wishes to examine the
structure of the code\footnote{The maintenance of two separate codes is
  primarily intended to provide an efficient way for the authors to
  confirm that each update is free from implementation errors.}. 

The library of subroutines, the command-line version and the online
version share a common set of features, which we will describe first. We
will then give operating instructions for each of these three 
{\tt HiggsBounds} formats individually. 

\subsection{Common features: Input}
\label{subsec:Common features: Input}
{\tt HiggsBounds} requires five types of input:

\begin{itemize}
\item the number of neutral Higgs bosons in the model under study ({\tt nHzero}) 
\item the number of singly, positively charged Higgs bosons in the model under study ({\tt nHplus}) 
\item the set of experimental analyses which should be considered
  ({\tt whichanalyses}) 
\item the theoretical predictions of the model under study (a set of
  input arrays) 
\item the format of these theoretical predictions ({\tt whichinput})
\end{itemize}

\refta{table:instructionsA} contains further information on the variable
{\tt nHzero} and {\tt nHplus}, and the possible values of {\tt whichanalyses} are described in
\refta{table:instructions1}. Note that the option {\tt whichanalyses=`singH'}
should only be used if neither processes involving Higgs pair production
at LEP nor processes involving the $h_j\to h_ih_i$ decay are
relevant. However, if these conditions are met, this option can save
significantly on computing time. 

{\tt HiggsBounds} expects the theoretical input to be in one of three
formats, labelled by the variable {\tt whichinput}. These formats were
described in detail in \refse{sec:inputdescription} and
are briefly summarised in \refta{table:instructions2}. In
\refta{table:instructions3}, \refta{table:instructionspart1b}, \refta{table:instructions3part2} and \refta{table:instructions3a} we assign names
to all of the possible input arrays (each array is defined in terms of
the notation used in \refse{sec:inputdescription}). These names will
prove useful when we describe the input requirements of each version of
{\tt HiggsBounds} individually. 

\begin{table}[!h]
\begin{tabular}{lll}
{\tt nHzero} & {\tt nHplus}\mystar & ({\tt integer}) \\
\hline
0-9& 0-9& This is the default range.\\
& & It can easily be extended by the user if required.
\end{tabular}\\
\caption{\sl\label{table:instructionsA}The possible values of the variable 
  {\tt nHzero}, which labels the number of neutral Higgs bosons 
in the model under
  study and the variable {\tt nHplus}, which labels the number of singly, positively charged Higgs.
} 
\end{table}

\begin{table}[!h]
\begin{tabular}{ll}
{\tt whichanalyses} & ({\tt character(LEN=5)}) \\
\hline
{\tt LandT} & both LEP and Tevatron analyses \\
{\tt onlyL} & only LEP analyses\\
{\tt onlyT} & only Tevatron analyses\\
{\tt onlyP}\mystar & only published analyses (defined as analyses with an arXiv number)\\
{\tt singH} & only analyses for processes involving one Higgs boson\\
\end{tabular}\\
\caption{\sl\label{table:instructions1}The possible values of the variable {\tt
    whichanalyses}, which indicates which subset of experimental analyses
  will be considered by {\tt HiggsBounds}.
} 
\end{table}

\begin{table}[!h]
\begin{tabular}{ll}
{\tt whichinput} & ({\tt character(LEN=4)}) \\
\hline
{\tt effC} & Masses, total decay widths, \\[-2mm]
           & ratios of effective couplings squared, some branching ratios.\\
{\tt part} & Masses, total decay widths, ratios of LEP cross sections, \\[-2mm]
           & mainly ratios of partonic Tevatron cross sections, branching ratios.\\
{\tt hadr} & Masses, total decay widths, ratios of LEP cross sections, \\[-2mm]
           & ratios of hadronic Tevatron cross sections, branching ratios.\\
\end{tabular}\\
\caption{\sl\label{table:instructions2}The possible values of the variable {\tt
    whichinput}, which indicates the format of the theoretical
  predictions provided by the user for the neutral Higgs sector. (See \refse{sec:inputdescription} for a more detailed description of each of
  these settings).} 
\end{table}

\clearpage

\begin{table}[h]
\begin{tabular}{llll}
input arrays &  \multicolumn{3}{l}{\tt (double precision)}\\
\hline
{\tt Mh(nHzero)}         &  $m_{h_i}$ &               \multicolumn{2}{l}{in GeV}\\
{\tt MhGammaTot(nHzero)} &  $\Gamma_{\rm tot}(h_i)$ & \multicolumn{2}{l}{in GeV} \\
\hline
{\tt Mhplus(nHplus)}\mystar         &  $m_{H_i^{\pm}}$ &               \multicolumn{2}{l}{in GeV}\\
{\tt MhplusGammaTot(nHplus)}\mystar &  $\Gamma_{\rm tot}(H_i^{\pm})$ & \multicolumn{2}{l}{in GeV} \\
\hline
{\tt g2hjss\_s(nHzero)}\mystar      & $\left(\frac{g^{\rm model}_{s,h_j{(\text{OP})}}}
                                    {g^{\rm SM}_{H{(\text{OP})}}}\right)^2$,& OP =& $s\bar{s}$     \\
{\tt g2hjcc\_s(nHzero)}\mystar      & & & $c\bar{c}$     \\
{\tt g2hjbb\_s(nHzero)}\mystar      & & & $b\bar{b}$     \\
{\tt g2hjtoptop\_s(nHzero)}\mystar      & & & $t\bar{t}$     \\
{\tt g2hjtautau\_s(nHzero)}\mystar  & & & $\tau^+\tau^-$ \\
{\tt g2hjss\_p(nHzero)}\mystar      & $\left(\frac{g^{\rm model}_{p,h_j{(\text{OP})}}}
                                    {g^{\rm SM}_{H{(\text{OP})}}}\right)^2$,& OP =& $s\bar{s}$     \\
{\tt g2hjcc\_p(nHzero)}\mystar      & & & $c\bar{c}$     \\
{\tt g2hjbb\_p(nHzero)}\mystar      & & & $b\bar{b}$     \\
{\tt g2hjtoptop\_p(nHzero)}\mystar      & & & $t\bar{t}$     \\
{\tt g2hjtautau\_p(nHzero)}\mystar & & & $\tau^+\tau^-$ \\
{\tt g2hjWW(nHzero)}      &$\left(\frac{g^{\rm model}_{h_j{(\text{OP})}}}
                                    {g^{\rm SM}_{H{(\text{OP})}}}\right)^2$,  
				& OP = & $W^+W^-$           \\
{\tt g2hjZZ(nHzero)}      & & & $ZZ$           \\
{\tt g2hjZga(nHzero)}\mystar     & & & $Z\gamma$      \\
{\tt g2hjgaga(nHzero)}    & & & $\gamma\gamma$ \\
{\tt g2hjgg(nHzero)}      & & & $gg$           \\
\hline
{\tt g2hjhiZ(nHzero,nHzero)} & $\left(\frac{g^{\rm model}_{h_jh_iZ}}
                                   {g^{\rm ref}_{HH^{\prime}Z}}\right)^2$ &&\\
\hline
\end{tabular}\\

\caption{\sl\label{table:instructions3}Input arrays for model predictions for 
	 effective normalised squared couplings 
   recognised by {\tt HiggsBounds}.
 The size of
 each array is given in brackets in the first column. See \refse{sec:inputdescription} for the description of the notation used
 in the second column.} 
\end{table}
\begin{table}[h]
\begin{tabular}{llll}
input arrays cont. &  \multicolumn{3}{l}{\tt (integer)}\\
\hline
{\tt CP\_value(nHzero)}\mystar,     & where {\tt CP\_value(i)}& = -1 & if $i$th neutral Higgs is CP-odd       \\
                         &                    & = 0  & if $i$th neutral Higgs has mixed CP         \\
                         &                    & = 1  & if  $i$th neutral Higgs is CP-even        \\
\hline
\end{tabular}\\

\caption{\sl\label{table:instructionspart1b}Input array to specify the CP
properties of each of the neutral Higgs bosons predicted by the model.
 The size of
 the array is given in brackets in the first column.} 
\end{table}

\begin{table}[h]
\begin{tabular}{llll}
input arrays cont.&  \multicolumn{3}{l}{\tt (double precision)}\\
\hline
{\tt BR\_hjss(nHzero)} \mystar      & BR($h_j \to {\text{OP}}$),& OP = &$s\bar{s}$     \\
{\tt BR\_hjcc(nHzero)} \mystar   & &                               &$c\bar{c}$     \\
{\tt BR\_hjbb(nHzero)}   & &                               &$b\bar{b}$     \\
{\tt BR\_hjtautau(nHzero)}& &                               &$\tau^+\tau^-$ \\
{\tt BR\_hjWW(nHzero)}    & &                               &$W^+W^-$           \\
{\tt BR\_hjZZ(nHzero)}    & &                               &$ZZ$           \\
{\tt BR\_hjZga(nHzero)} \mystar  & &                               &$Z\gamma$      \\
{\tt BR\_hjgaga(nHzero)}  & &                               &$\gamma\gamma$ \\
{\tt BR\_hjgg(nHzero)}  \mystar  & &                               &$gg$           \\
\hline
{\tt BR\_hjinvisible(nHzero)} \mystar   &  \multicolumn{3}{l}{BR($h_j\to {\rm invisible}$)}  \\
{\tt BR\_hjhihi(nHzero,nHzero)} & \multicolumn{3}{l}{BR($h_j\to h_ih_i$)} \\
\hline
{\tt BR\_tWpb}\mystar& BR($t \to W^{+}b$)&    \\
{\tt BR\_tHpjb(nHplus)}\mystar& BR($t \to H_j^{+}b$)&    \\
\hline
{\tt BR\_Hpjcs(nHplus)}\mystar& BR($H_j^{+} \to {\text{OP}}$),& OP = &$c\bar{s}$     \\
{\tt BR\_Hpjcb(nHplus)}\mystar& & &$c\bar{b}$     \\
{\tt BR\_Hpjtaunu(nHplus)}\mystar& &  &$\tau^+ \bar\nu_{\tau}$     \\
\hline
\end{tabular}\\
\caption{\sl\label{table:instructions3part2}Input arrays for model predictions for branching ratios
    recognised by {\tt HiggsBounds}.
  The size of
  each array is given in brackets in the first column. See \refse{sec:inputdescription} for the description of the notation used
  in the second column.
  The elements of {\tt BR\_hjhihi} are ordered such that 
  {\tt BR\_hjhihi(j,i)}$ = \BR(h_j\to h_ih_i)$. }
\end{table}

\begin{table}[!h]
\begin{tabular}{llll}
input arrays cont. & \multicolumn{3}{l}{\tt (double precision)}\\
\hline 
{\tt CS\_lep\_hjZ\_ratio(nHzero)}             & $R_{\sigma}(P)$,
                                          & $P$ =
                                            & $e^+e^- \to h_j Z$   \\
{\tt CS\_lep\_bbhj\_ratio(nHzero,nHzero)} \mystar        & & & $e^+e^- \to b \bar{b} h_j$ \\
{\tt CS\_lep\_tautauhj\_ratio(nHzero,nHzero)} \mystar        & & & $e^+e^- \to \tau^+ \tau^- h_j$ \\
{\tt CS\_lep\_hjhi\_ratio(nHzero,nHzero)}         & & & $e^+e^- \to h_j h_i$ \\
\hline
{\tt CS\_lep\_HpjHmj\_ratio(nHzero)}\mystar & & & $e^+e^- \to H^+_j H^-_j$ \\
\hline
{\tt CS\_tev\_pp\_hj\_ratio(nHzero)}           & & & $p\bar{p} \to h_j$   \\
{\tt CS\_tev\_pp\_hjb\_ratio(nHzero)}          & & & $p\bar{p} \to b h_j$ \\
{\tt CS\_tev\_pp\_hjW\_ratio(nHzero)}          & & & $p\bar{p} \to h_j W$ \\
{\tt CS\_tev\_pp\_hjZ\_ratio(nHzero)}          & & & $p\bar{p} \to h_j Z$\\
{\tt CS\_tev\_pp\_vbf\_ratio(nHzero)}          & & & $p\bar{p} \to h_j 
                                                 {\rm \, via \, VBF}$\\
{\tt CS\_tev\_pp\_tthj\_ratio(nHzero)}\mystar  & & & $p\bar{p} \to t \bar{t} h_j$ \\
\hline
{\tt CS\_tev\_gg\_hj\_ratio(nHzero)}           & $R^{h_j}_{nm}$,
                                          & $nm$ = 
                                            & $gg$ \\
{\tt CS\_tev\_bb\_hj\_ratio(nHzero)}           & & & $b\bar{b}$   \\
\hline
{\tt CS\_tev\_ud\_hjWp\_ratio(nHzero)}         & $R^{h_j +W^+}_{nm}$,
                                          & $nm$ = 
                                            & $u\bar{d}$ \\
{\tt CS\_tev\_cs\_hjWp\_ratio(nHzero)}         & & & $c\bar{s}$ \\
\hline
{\tt CS\_tev\_ud\_hjWm\_ratio(nHzero)}         &$R^{h_j +W^-}_{nm}$
                                          & $nm$ =
                                            &  $d\bar{u}$ \\
{\tt CS\_tev\_cs\_hjWm\_ratio(nHzero)}         & & &  $s\bar{c}$ \\
\hline
{\tt CS\_tev\_dd\_hjZ\_ratio(nHzero)}          &$R^{h_j +Z}_{nm}$
                                          & $nm$ =                
                                            & $d\bar{d}$ \\
{\tt CS\_tev\_uu\_hjZ\_ratio(nHzero)}          & & & $u\bar{u}$ \\
{\tt CS\_tev\_ss\_hjZ\_ratio(nHzero)}          & & & $s\bar{s}$ \\
{\tt CS\_tev\_cc\_hjZ\_ratio(nHzero)}          & & & $c\bar{c}$ \\
{\tt CS\_tev\_bb\_hjZ\_ratio(nHzero)}          & & & $b\bar{b}$ \\
\hline
{\tt CS\_tev\_bg\_hjb\_ratio(nHzero)}          &$R^{h_j +b,h_j +\bar{b}}_{nm}$
                                          & $nm$ = 
                                            &  $bg,\bar{b}g$ \\
\hline
\end{tabular}\\
\caption{\sl\label{table:instructions3a}Input arrays 
	for model predictions for cross section ratios
	recognised by {\tt HiggsBounds}.
  The size of
  each array is given in brackets in the first column. The LEP or
  hadronic Tevatron cross section ratios $R_{\sigma}(P)$ are defined
\refeq{eq:R-sigma}, and the partonic Tevatron cross section ratios
  $R^{h_j+y}_{nm}$ are defined in \refeq{eq:R-nm-partonic}.
\vspace*{5mm}} 
\end{table}

\clearpage

\subsection{Common features: Output}

{\tt HiggsBounds} provides the user with four types of output:

\begin{itemize}
\item whether the parameter point is excluded at the 95\% C.L. or not
  ({\tt HBresult}) 
\item the reference number of the 
analysis application ($X_0$)
with the highest statistical sensitivity ({\tt chan}).  
\item the number of Higgs bosons which have contributed to the theoretical rate
  for the corresponding process ({\tt ncombined}) 
\item the ratio of the theoretical rate $Q_{\rm model}$ to the observed
  limit $Q_{\rm obs}$ for this process ({\tt obsratio}). 
\end{itemize}

\refta{table:instructions4} shows the possible values of {\tt HBresult}
and {\tt obsratio}, which are complementary. \refta{table:instructions4a}
and \refta{table:instructions4b} gives information on 
{\tt chan} and {\tt ncombined}
respectively. If the library of subroutines or the command-line versions
are used, the key 
associating the
reference numbers with the analysis applications
is written in the file {\tt
  Key.dat}. In the online version, this information appears on the
screen.  

\begin{table}[!h]
\begin{tabular}{lll}
\hline
{\tt HBresult} &{\tt obsratio}&\\
 ({\tt integer})&({\tt double precision})&\\
\hline
           0 &   $\ge 1.0$   & parameter point is excluded\\
           1 &   $<1.0$      & parameter point is not excluded\\
          -1 &   $\le 0.0$   & invalid parameter set\\
\end{tabular}
\caption{\sl\label{table:instructions4}The possible values of the output variables {\tt
    HBresult} and {\tt obsratio}, which indicate whether a parameter
  point has been excluded at the 95\% C.L. by the experimental results
  under consideration.
\vspace*{5mm}} 
\end{table} 

\begin{table}[!h]
\begin{tabular}{ll}
{\tt chan} & ({\tt integer})\\
\hline
0-[\# of considered analyses]
 & See the file {\tt Key.dat} for the definition of each  \\[-2mm]
 & reference number. {\tt Key.dat} is automatically\\ [-2mm]
 & generated when either the command line or the\\[-2mm]
 & subroutine version of {\tt HiggsBounds} are used.
\end{tabular}\\
\caption{\sl\label{table:instructions4a}
  Further information about the output variable
  {\tt chan}, which stores the reference number of the 
  analysis application
  with the
  highest statistical sensitivity. (0= no process applies)
\vspace*{5mm}} 
\end{table} 

\begin{table}[!h]
\begin{tabular}{ll}
{\tt ncombined}& ({\tt integer})\\
\hline
1-{\tt nHzero}
	& Number of neutral (or singly, positively charged) Higgs bosons\\[-2mm] 
                     & which have contributed to the theoretical rate for this process. 
\end{tabular}\\
\caption{\sl\label{table:instructions4b}Further information 
about the output variable {\tt ncombined}.
\vspace*{5mm}}
\end{table}

\subsection{Library of subroutines}

\subsubsection*{Installation}

The {\tt HiggsBounds} code can be compiled to form a library of
subroutines using the following commands: 

\begin{verbatim}
./configure
make libHB
\end{verbatim}

A program for which the {\tt HiggsBounds} subroutines should be used
can be
compiled and linked to the library by adding {\tt -L<HBpath> -lHB} to
the command line, for example, 

\begin{verbatim}
gfortran myprog.f90 -o myprog -L<HBpath> -lHB
\end{verbatim}

where {\tt <HBpath>} is the location of the {\tt HiggsBounds} library. 

The {\tt HiggsBounds} subroutines make use of the file handles 
10, 11, 44, 45 and 87 
(i.e. users should avoid using these file handles in the program which calls these subroutines.).

\subsubsection*{Subroutine {\tt initialize\_HiggsBounds}\mystarbrac}

The subroutine {\tt initialize\_HiggsBounds} must be called before any
other {\tt Higgs\-Bounds} subroutine. It performs some preparatory
operations such as reading in the tables of data. It is called as: 

\begin{verbatim}
call initialize_HiggsBounds(nHzero, nHplus, whichanalyses)
\end{verbatim}

When using the 
subroutines in another code, the subroutine
{\tt initialize\_HiggsBounds} must be called only once,
before any other {\tt Higgs\-Bounds} subroutine is called. 

If the user does not wish to test the neutral Higgs sector with 
{\tt HiggsBounds}, then he/she should set {\tt nHzero=0}.  Similarly, if the user
does not wish to test the charged Higgs sector with {\tt HiggsBounds}, then
he/she should set {\tt nHplus=0}.
\medskip

\subsubsection*{Subroutines {\tt HiggsBounds\_neutral\_input\_effC}\mystar,\\ 
{\tt HiggsBounds\_neutral\_input\_part}\mystar
and {\tt HiggsBounds\_neutral\_input\_hadr}\mystar}

The neutral Higgs sector input can be passed to {\tt HiggsBounds} using one 
of the 
subroutines {\tt HiggsBounds\_neutral\_input\_effC}, 
{\tt HiggsBounds\_neutral\_input\_part} or
{\tt HiggsBounds\_neutral\_input\_hadr}.
They set the value of {\tt whichinput} to be {\tt effC},
{\tt part} and {\tt hadr} respectively and therefore require different
arguments.  The use of one of these subroutines is only required if {\tt
nHzero} is non-zero (recall that {\tt nHzero} is set in subroutine {\tt
initialize\_HiggsBounds}).  These subroutines are called as:
\begin{verbatim}
call HiggsBounds_neutral_input_effC(Mh,MhGammaTot,
     &  g2hjss_s,g2hjss_p,g2hjcc_s,g2hjcc_p,g2hjbb_s,g2hjbb_p,
     &  g2hjtoptop_s,g2hjtoptop_p,g2hjtautau_s,g2hjtautau_p,            
     &  g2hjWW,g2hjZZ,g2hjZga,g2hjgaga,g2hjgg,                     
     &  g2hjhiZ,BR_hjinvisible,BR_hjhihi                  )
\end{verbatim}

\begin{verbatim}
call HiggsBounds_neutral_input_part(Mh,MhGammaTot, CP_value,                   
     &  CS_lep_hjZ_ratio,CS_lep_bbhj_ratio,CS_lep_tautauhj_ratio,                
     &  CS_lep_hjhi_ratio,                            
     &  CS_tev_gg_hj_ratio,  CS_tev_bb_hj_ratio,        
     &  CS_tev_bg_hjb_ratio,                          
     &  CS_tev_ud_hjWp_ratio,CS_tev_cs_hjWp_ratio,     
     &  CS_tev_ud_hjWm_ratio,CS_tev_cs_hjWm_ratio,     
     &  CS_tev_dd_hjZ_ratio, CS_tev_uu_hjZ_ratio,      
     &  CS_tev_ss_hjZ_ratio, CS_tev_cc_hjZ_ratio,      
     &  CS_tev_bb_hjZ_ratio,                          
     &  CS_tev_pp_vbf_ratio, CS_tev_pp_tthj_ratio,                         
     &  BR_hjss,BR_hjcc,BR_hjbb,BR_hjtautau,                          
     &  BR_hjWW,BR_hjZZ,BR_hjZga,BR_hjgaga,BR_hjgg,BR_hjinvisible,                             
     &  BR_hjhihi                                    )
\end{verbatim}

\begin{verbatim}
call HiggsBounds_neutral_input_hadr(Mh,MhGammaTot, CP_value,                  
     &  CS_lep_hjZ_ratio,CS_lep_bbhj_ratio,CS_lep_tautauhj_ratio,                 
     &  CS_lep_hjhi_ratio,           
     &  CS_tev_pp_hj_ratio,  CS_tev_pp_hjb_ratio,      
     &  CS_tev_pp_hjW_ratio, CS_tev_pp_hjZ_ratio,      
     &  CS_tev_pp_vbf_ratio, CS_tev_pp_tthj_ratio,                        
     &  BR_hjss,BR_hjcc,BR_hjbb,BR_hjtautau,                          
     &  BR_hjWW,BR_hjZZ,BR_hjZga,BR_hjgaga,BR_hjgg,BR_hjinvisible,                          
     &  BR_hjhihi                                    )
\end{verbatim}

Each of these arguments must be supplied. However, if 
a branching ratio, effective coupling or cross section is believed 
to be irrelevant, the corresponding array may be filled with zeros. 
This will ensure that the value of
$Q_{\rm model}$ for processes involving this quantity will also be
zero. For example, in the MSSM, scenarios exist where the decay $h_j \to
\gamma \gamma$ will certainly not be the process with the highest statistical
sensitivity of $Q_{\rm model}/Q_{\rm expec}$ and, consequently, 
it may be convenient to set the arrays
{\tt g2hjgaga} and {\tt BR\_hjgaga} to zero for simplicity in such a
case. 
Note that if a quantity used in a SM-likeness test is set to zero, the model
point will fail that SM-likeness test, even if the corresponding SM quantity
is very small.

Also, depending on the value given for {\tt whichanalyses}, some of the
input arrays will be ignored within {\tt HiggsBounds}. For example, if
{\tt whichanalyses=`onlyT'}, the branching ratio for the Higgs cascade decay
$h_j \to h_i h_i$
will not be relevant. Therefore, setting this array to zero
will not affect the {\tt HiggsBounds} results in this case. 
\medskip

\subsubsection*{Subroutine {\tt HiggsBounds\_charged\_input}\mystar}

The subroutine {\tt HiggsBounds\_charged\_input} gives the charged Higgs
sector input to HiggsBounds.  The use of this subroutine is only required if
{\tt nHplus} is non-zero (recall that {\tt nHplus} is set in subroutine {\tt
initialize\_HiggsBounds}).  It is called as:
\begin{verbatim}
call HiggsBounds_charged_input(MHplus, MHplusGammaTot
     &  CS_lep_HpjHmi_ratio,
     &  BR_tWpb,BR_tHpjb,BR_Hpjcs,BR_Hpjcb,BR_Hptaunu       )
\end{verbatim}

\subsubsection*{Subroutine {\tt run\_HiggsBounds}\mystar}

This subroutine performs the main part of the {\tt
HiggsBounds} calculations. It is called as:
\begin{verbatim}
call run_HiggsBounds(  HBresult,chan,                          
     &                 obsratio, ncombined      )
\end{verbatim}

\subsubsection*{Subroutine {\tt finish\_HiggsBounds}}

The subroutine {\tt finish\_HiggsBounds} should be called once at the
end of the program, after all other {\tt HiggsBounds}
subroutines\footnote{In the Fortran 90 version of the code, the
  subroutine {\tt finish\_HiggsBounds} is used to deallocate the
  allocatable arrays used within {\tt HiggsBounds}.}. It is called as: 
 
\begin{verbatim}
call finish_HiggsBounds
\end{verbatim}

\subsubsection*{Functions for SM branching ratios, total decay width
  and cross sections}

The {\tt HiggsBounds} library also allows users access to the SM
Higgs branching ratios, total decay width and production cross sections,
which are used internally by {\tt HiggsBounds}. We use SM Higgs
branching ratios and total decay width from the program 
HDECAY 3.4 \citehdecay. 
The SM hadronic cross sections have been obtained from 
the TEV4LHC Higgs Working Group \cite{hep-ph/0612172}
(see \refta{table:SMfunctions} for references to the original works)
with the exception of the $\sigma^{\rm SM}(p\bar{p}\to b g \to b H)$ and $\sigma^{\rm SM}(p\bar{p}\to \bar{b} g \to \bar{b} H)$
cross sections. The latter cross sections have been calculated with
the program HJET 1.3 \cite{hep-ph/0305321,arXiv:0705.2744,arXiv:1003.4438} 
for a set of different cuts
on the transverse momentum and pseudo-rapidity of the $b$-quark,
which are needed internally in order to apply correctly the results
of some Tevatron analyses and which were not available from
\cite{hep-ph/0612172}. From this set, only the cross section 
without cuts is externally provided.

Also included is a function for the SM top quark decay width into $W^+ b$ in $\gev$, which depends
 on the top quark pole mass in $\gev$, given at next-to-leading-order, neglecting terms of order
 $m_b^2/m_t^2$, $\alpha_s^2$ and $\left(\alpha_s/\pi\right)M_W^2/m_t^2$, 
as quoted in \cite{Amsler:2008zzb} (original reference: \cite{Jezabek:1988iv}).
 This is not used within {\tt HiggsBounds}, but may prove useful when calculating $\BR(t\to H^+b)$.

Note that these functions are only valid within certain mass ranges, which
are sufficient for the requirements of {\tt HiggsBounds}.
 If a function is called with an argument outside this range, it returns a
value of $-1$.

\medskip

\begin{table}
\begin{tabular}{lllll}
\hline
        function                &\multicolumn{3}{l}{({\tt double precision})} & \\
\hline
      {\tt  SMGamma\_h(Mh)     } &\multicolumn{3}{l}{$\Gamma^{\rm SM}_{\rm tot} (h_i)$}&\citehdecay
\\
      {\tt  SMBR\_Hss(Mh)      } & ${\rm BR}^{\rm SM}$($H \to$ OP), & OP= & $s \bar{s}$&\citehdecay
\\
      {\tt  SMBR\_Hcc(Mh)      } &                      &     & $c \bar{c}$&\citehdecay
\\
      {\tt  SMBR\_Hbb(Mh)      } &                      &     & $b \bar{b}$&\citehdecay
\\
      {\tt  SMBR\_Htoptop(Mh)  } &                      &     & $t \bar{t}$&\citehdecay
\\
      {\tt  SMBR\_Htautau(Mh)  } &                      &     & $\tau^- \tau^+ $&\citehdecay
\\
      {\tt  SMBR\_Hmumu(Mh)    } &                      &     & $\mu^- \mu^+   $&\citehdecay
\\
      {\tt  SMBR\_HWW(Mh)      } &                      &     & $W^+W^-            $&\citehdecay
\\
      {\tt  SMBR\_HZZ(Mh)      } &                      &     & $ZZ            $&\citehdecay
\\
      {\tt  SMBR\_HZgam(Mh)    } &                      &     & $Z \gamma      $&\citehdecay
\\
      {\tt  SMBR\_Hgamgam(Mh)  } &                      &     & $\gamma \gamma $&\citehdecay
\\
      {\tt  SMBR\_Hgg(Mh)      } &                      &     & $gg            $&\citehdecay
\\
\hline
      {\tt  SMGamma\_tWpb(Mtop)}\mystar &  ${\rm \Gamma}^{\rm SM}$($t \to W^+ b$ )&     &                 & \cite{Jezabek:1988iv}\\
\hline
      {\tt  SMCS\_tev\_pp\_qq\_HW(Mh)} & $\sigma^{\rm SM}(P)$, & P=  & $p\bar{p}\to q\bar{q}\to HW$&\citeSMVHCS
\\
      {\tt  SMCS\_tev\_pp\_qq\_HZ(Mh)} &                      &     & $p\bar{p}\to q\bar{q}\to HZ$&\citeSMVHCS
\\
      {\tt  SMCS\_tev\_pp\_gg\_H(Mh) } &                      &     & $p\bar{p}\to gg      \to H $&\citeSMgluonfusionCS
\\
      {\tt  SMCS\_tev\_pp\_bb\_H(Mh) } &                      &     & $p\bar{p}\to b\bar{b}\to H $&\cite{hep-ph/0304035}
\\
      {\tt  SMCS\_tev\_pp\_vbf\_H(Mh)} &                      &     & $p\bar{p}\to H $ via VBF &\citeSMVBFCS
\\
      {\tt  SMCS\_tev\_pp\_ttH(Mh)}\mystar    &                      &     & $p\bar{p}\to t\bar{t}H  $&\citeSMttHCS \\
      {\tt  SMCS\_tev\_pp\_bg\_Hb(Mh)} 
	&\multicolumn{3}{l}{  $\sigma^{\text{SM}}(p\bar{p}\to bg\to Hb) 
		+\sigma^{\text{SM}}(p\bar{p}\to \bar{b}g\to H\bar{b}) $} &\cite{hep-ph/0305321,arXiv:0705.2744,arXiv:1003.4438}
\\
\hline
\end{tabular}\\
\caption{\sl\label{table:SMfunctions}
  SM Higgs branching ratios, total decay widths in units of GeV
  and hadronic Tevatron cross sections in units of pb
  provided as functions by
  {\tt HiggsBounds}, together with references. Each function takes a
  Higgs mass {\tt Mh (double precision)} as its argument. Also included is the SM top quark decay width
 into $W^+ b$ in $\gev$, which depends on the top quark pole mass in $\gev$.
 If a function is called with an argument outside its mass range, it returns a value of -1.
\vspace*{5mm}} 
\end{table}

\subsubsection*{Examples}

We have provided three example programs which demonstrate the use of the
{\tt Higgs\-Bounds} subroutines. The first example relates to
the Fourth Generation Model and is contained in the file {\tt
   example-SM\_vs\_4thGen.F}. 
This program uses the {\tt HiggsBounds} functions for
the SM branching ratios and SM total decay width to calculate the Higgs
decay width and the effective normalised squared couplings
in the SM and a simple Fourth Generation Model. 
This information is then used as input for
the subroutine {\tt HiggsBounds\_neutral\_input\_effC}, which is called
once with SM input and once with Fourth Generation Model input.
Once the {\tt HiggsBounds} library has been compiled
(using {\tt ./configure ; make libHB} as described previously), the code
 {\tt example-SM\_vs\_4thGen.F} can be compiled and run with the commands: 

\begin{verbatim}
gfortran example-SM_vs_4thGen.F -o example-SM_vs_4thGen \
          -L<HBpath> -lHB
./example-SM_vs_4thGen
\end{verbatim}

where {\tt <HBpath>} is the location of the {\tt HiggsBounds} library. 

The files {\tt HBwithFH.F} and {\tt HBwithCPsuperH.f} demonstrate the
use of 
the subroutine version of 
{\tt Higgs\-Bounds} with the publicly
available programs \feynhiggs\ and \cpsuperh, respectively. 
We refer the reader to the
extensive comments contained within these example files for further
details.

\subsection{Command line version}
\subsubsection*{Installation}

In order to be able to call {\tt HiggsBounds} from the command line, it
should be compiled using the commands 
\begin{verbatim}
./configure
make
\end{verbatim}

\subsubsection*{Command line and input file format\mystarbrac}

In the command-line usage of {\tt HiggsBounds}, the arrays containing
the theoretical model predictions are read from text files. The other
options are specified in the command line, which is of the form: 
\begin{verbatim}
./HiggsBounds <whichanalyses> <whichinput> <nHzero> <nHplus> <prefix>
\end{verbatim}

The variable {\tt <prefix>} is a string which is added to the front of
input and output file names and may include directory names or other
identifying information. 

\refta{table:contentsoffiles1} and \refta{table:contentsoffiles2} describe
the contents of each input file.  Note that each input file should start
with a line number.  The input files should not contain any comments or
blank lines.  The line number identifies the predictions which belong to the
same model parameter point in different files.  
The file {\tt BR\_H\_OP.dat}\mystar\ 
contains neutral Higgs branching ratios with a SM equivalent (i.e. 
`OP'=`ordinary particles'), whereas 
the file {\tt BR\_H\_NP.dat}\mystar\ contains
neutral Higgs branching ratios without a SM equivalent (i.e.  `NP'=`new
particles')\footnote{In previous versions of {\tt HiggsBounds}, this was
equivalent to dividing the neutral Higgs branching ratios between a file
containing one type of neutral Higgs and a file containing two types of neutral
Higgs bosons.  However, the input to {\tt HiggsBounds} 2.0.0 requires the
branching ratio of the Higgs to invisible particles, hence the new
filenames.}.

\begin{table}[!t]
\begin{tabular}{ll}
\hline
file name & data format\\
\hline
{\tt MH\_GammaTot.dat}                & {\tt k},          {\tt Mh, MhGammaTot}                                 \\[-1mm]
{\tt MHplus\_GammaTot.dat}\mystar    & {\tt k},          {\tt Mhplus, MhplusGammaTot}                         \\[-1mm]
{\tt effC.dat}  \mystarbrac             & {\tt k},          {\tt g2hjss\_s,g2hjss\_p,g2hjcc\_s,g2hjcc\_p}                       \\[-1mm]  
                                      &\phantom{{\tt k},} {\tt g2hjbb\_s,g2hjbb\_p,g2hjtoptop\_s,g2hjtoptop\_p}          \\[-1mm]  
                                      &\phantom{{\tt k},} {\tt g2hjtautau\_s,g2hjtautau\_p,}          \\[-1mm]  
                                      &\phantom{{\tt k},} {\tt g2hjWW,g2hjZZ,g2hjZga,}               \\[-1mm]  
                                      &\phantom{{\tt k},} {\tt g2hjgaga,g2hjgg,}        \\[-2mm]
                                      &\phantom{{\tt k},} {some elements of {\tt g2hjhiZ}}            \\[-2mm]
                                      &\phantom{{\tt k},} {(lower left triangle - see example)}       \\[-1mm]    
{\tt LEP\_HZ\_CS\_ratios.dat}         & {\tt k},          {\tt CS\_lep\_hjZ\_ratio}                            \\[-1mm]  
{\tt LEP\_H\_ff\_CS\_ratios.dat}\mystar& {\tt k},          {\tt CS\_lep\_bbhj\_ratio}, {\tt CS\_lep\_tautauhj\_ratio}                        \\[-1mm]     
{\tt LEP\_2H\_CS\_ratios.dat}         & {\tt k},          some elements of {\tt CS\_lep\_hjhi\_ratio} \\[-2mm]
                                      &\phantom{{\tt k},} (lower left triangle - see example)         \\[-1mm]
{\tt LEP\_HpHm\_CS\_ratios.dat}\mystar& {\tt k},          {\tt CS\_lep\_HpjHmj\_ratio}  \\[-2mm]
{\tt TEV\_H\_0jet\_partCS\_ratios.dat}& {\tt k},          {\tt CS\_tev\_gg\_hj\_ratio,CS\_tev\_bb\_hj\_ratio}  \\[-1mm]   
{\tt TEV\_H\_1jet\_partCS\_ratios.dat}& {\tt k}           {\tt CS\_tev\_bg\_hjb\_ratio}                         \\[-1mm]  
{\tt TEV\_HW\_partCS\_ratios.dat}     & {\tt k},          {\tt CS\_tev\_ud\_hjWp\_ratio},{\tt CS\_tev\_cs\_hjWp\_ratio},             \\[-2mm]  
                                      &\phantom{{\tt k},} {\tt CS\_tev\_ud\_hjWm\_ratio},{\tt CS\_tev\_cs\_hjWm\_ratio}              \\[-1mm]  
{\tt TEV\_HZ\_partCS\_ratios.dat}     & {\tt k},          {\tt CS\_tev\_dd\_hjZ\_ratio}, {\tt CS\_tev\_uu\_hjZ\_ratio,}              \\[-2mm] 
                                      &\phantom{{\tt k},} {\tt CS\_tev\_ss\_hjZ\_ratio}, {\tt CS\_tev\_cc\_hjZ\_ratio,}              \\[-2mm] 
                                      &\phantom{{\tt k},} {\tt CS\_tev\_bb\_hjZ\_ratio}               \\[-1mm]
{\tt TEV\_H\_vbf\_hadCS\_ratios.dat}  & {\tt k},          {\tt CS\_tev\_pp\_vbf\_ratio}                        \\[-1mm] 
{\tt TEV\_H\_tt\_hadCS\_ratios.dat}\mystar   & {\tt k},          {\tt CS\_tev\_pp\_tthj\_ratio}                        \\[-1mm] 
{\tt TEV\_1H\_hadCS\_ratios.dat}\mystarbrac      & {\tt k},          {\tt CS\_tev\_pp\_hj\_ratio}, {\tt CS\_tev\_pp\_hjb\_ratio},              \\[-2mm]
                                      &\phantom{{\tt k},} {\tt CS\_tev\_pp\_hjW\_ratio},{\tt CS\_tev\_pp\_hjZ\_ratio,}              \\[-2mm]
                                      &\phantom{{\tt k},} {\tt CS\_tev\_pp\_vbf\_ratio}               \\[-1mm] 
                                      &\phantom{{\tt k},} {\tt CS\_tev\_pp\_tthj\_ratio}              \\[-1mm] 
\hline
\end{tabular}\\
\caption{\sl\label{table:contentsoffiles1}Names and data format of 
  all {\tt HiggsBounds} input files (part I).
  The right column shows the order of the input data arrays 
  within one line of the input file. For the order
  within the arrays, see example. 
  {\tt k} is the line number. 
  Note that the
  arrays {\tt CS\_tev\_pp\_vbf\_ratio}, {\tt CS\_tev\_pp\_tthj\_ratio} each
  appear in two different input files.  However, these files will never be
  required by {\tt HiggsBounds} simultaneously.
\vspace*{5mm}} 
\end{table}

\begin{table}[!t]
\begin{tabular}{ll}
\hline
file name & data format\\
\hline
{\tt BR\_H\_OP.dat}\mystar                      & {\tt k},          {\tt BR\_hjss,} \\[-1mm] 
                                      &\phantom{{\tt k},} {\tt BR\_hjcc,BR\_hjbb,BR\_hjtautau,}                  \\[-1mm] 
                                      &\phantom{{\tt k},} {\tt BR\_hjWW,BR\_hjZZ,BR\_hjZga},                               \\[-1mm] 
                                      &\phantom{{\tt k},} {\tt BR\_hjgaga,BR\_hjgg}               \\[-1mm] 
{\tt BR\_H\_NP.dat}\mystar                     & {\tt k},   {\tt BR\_hjinvisible},       some elements of {\tt BR\_hjhihi}                               \\[-2mm]
                                      &\phantom{{\tt k},} (row by row, without diagonal                          \\[-2mm]
                                      &\phantom{{\tt k},} - see example)                                         \\[-1mm]
{\tt BR\_t.dat}\mystar                & {\tt k}, {\tt BR\_tWpb}, {\tt BR\_tHpb}                              \\[-1mm]      
{\tt BR\_Hplus.dat}\mystar            & {\tt k}, {\tt BR\_Hpcs, BR\_Hpcb, BR\_Hptaunu  }                              \\[-1mm] 
{\tt CP\_values.dat}\mystar           & {\tt k}, {\tt CP\_value} \\
{\tt additional.dat}(optional)        & {\tt k},  .. . \\
\hline
\end{tabular}\\
\caption{\sl\label{table:contentsoffiles2}Names and data format of 
  all {\tt HiggsBounds} input files (part II).
  The right column shows the order of the input data arrays 
  within one line of the input file. For the order
  within the arrays, see example. 
  {\tt k} is the line number. Note that the
  arrays {\tt CS\_tev\_pp\_vbf\_ratio}, {\tt CS\_tev\_pp\_tthj\_ratio} each appear in two different input
  files. However, these files will never be required by 
  {\tt HiggsBounds} simultaneously.
\vspace*{5mm}} 
\end{table}

Care should be taken with the order of the array elements in the
files. This is best illustrated by an example, where we will use
$n_H=3$. The one dimensional arrays, e.g. {\tt Mh}, should be given in
the order 
\begin{quote}
{\tt Mh(1), Mh(2), Mh(3)}.
\end{quote}
\pagebreak
\clearpage
\pagebreak
However, not all of the elements of the two dimensional arrays are
required. Only the lower left triangle (including the diagonal) is
required from the arrays {\tt g2hjhiZ} and 
{\tt lepCS\_hjhi\_ratio},
since they are symmetric, e.g.  
\begin{align}
\begin{pmatrix}
{\tt g2hjhiZ(1,1)} & \Gray{\tt g2hjhiZ(1,2)} & \Gray{\tt g2hjhiZ(1,3)}\non\\
{\tt g2hjhiZ(2,1)} &      {\tt g2hjhiZ(2,2)} & \Gray{\tt g2hjhiZ(2,3)}\non\\
{\tt g2hjhiZ(3,1)} &      {\tt g2hjhiZ(3,2)} &      {\tt g2hjhiZ(3,3)}\non
\end{pmatrix}
\end{align}
i.e. the elements in the input file should be written in the order 
\begin{quote}
{\tt g2hjhiZ(1,1)}, {\tt g2hjhiZ(2,1)}, {\tt g2hjhiZ(2,2)}, 
{\tt g2hjhiZ(3,1)}, \\* {\tt g2hjhiZ(3,2)}, {\tt g2hjhiZ(3,3)} .
\end{quote}

For the array {\tt BR\_hjhihi}, only the off-diagonal components are required
\begin{align}
\begin{pmatrix} 
\Gray{\tt BR\_hjhihi(1,1)} & {\tt BR\_hjhihi(1,2)}     & {\tt BR\_hjhihi(1,3)} 
                                                                         \non\\
{\tt BR\_hjhihi(2,1)}      & \Gray{\tt BR\_hjhihi(2,2)} & {\tt BR\_hjhihi(2,3)} 
                                                                          \non\\
{\tt BR\_hjhihi(3,1)}      & {\tt BR\_hjhihi(3,2)}      & 
                                                \Gray{\tt BR\_hjhihi(3,3)}\non
\end{pmatrix}
\end{align}
since the diagonal elements are not physical quantities. Therefore, the
elements should be written in the order  
\begin{quote}
{\tt BR\_hjhihi(1,2)}, {\tt BR\_hjhihi(1,3)}, {\tt BR\_hjhihi(2,1)},
{\tt BR\_hjhihi(2,3)}, \\* {\tt BR\_hjhihi(3,1)}, {\tt BR\_hjhihi(3,2)}  
\end{quote}
in the input file.

The file {\tt additional.dat} is optional. If it is included, it can
have any number of columns greater than 1 (as for the previous files,
the first entry on each line should be the line number). It is envisaged
that this input file will be particularly useful when parameter scans
are performed over a variable which is not required by {\tt HiggsBounds}
but helpful when plotting the results.  For example, in the case of the MSSM, {\tt additional.dat} could be used to store the values of
$\tan \beta$. 

As in the subroutine version, the command line version of 
{\tt HiggsBounds} expects a subset of the total list of input arrays, which
depends on the chosen setting of {\tt whichinput}. The maximal list of files used for
each value of {\tt whichinput} is given in \refta{table:instructions1}.  

As discussed for the subroutine version, some of the arrays will not be
relevant for some of the choices for {\tt whichanalyses}. The command line
version of {\tt HiggsBounds} will consider the list of input files
appropriate to the setting {\tt whichinput} and then only attempt to
read any of these input files if the value chosen for {\tt whichanalyses}
means that at least one of the arrays it contains will be directly
used. \refta{table:instructions6} contains a list of which input files are
actually relevant to each value of {\tt whichanalyses}.  
For example, if {\tt whichinput = 'hadr'}, {\tt whichanalyses = 'LandT'}, {\tt nHzero > 0} and {\tt nHplus > 0}
then {\tt HiggsBounds} requires the input files: 
\begin{quote}
{\tt MH\_GammaTot.dat}, {\tt MHplus\_GammaTot.dat}, {\tt CP\_values.dat},\\* 
{\tt BR\_H\_NP.dat}, {\tt BR\_H\_OP.dat}, {\tt BR\_t.dat}, {\tt BR\_Hplus.dat},\\ 
{\tt LEP\_HZ\_CS\_ratios.dat}, {\tt LEP\_H\_ff\_CS\_ratios.dat},
{\tt LEP\_2H\_CS\_ratios.dat},\\* 
{\tt LEP\_HpHm\_CS\_ratios.dat}, {\tt TEV\_1H\_hadCS\_ratios.dat} .
\end{quote}

However, if {\tt whichinput = 'hadr'}, {\tt whichanalyses = 'onlyL'}, {\tt nHzero > 0} and {\tt nHplus > 0} 
  {\tt HiggsBounds} requires the input files: 
\begin{quote}
{\tt MH\_GammaTot.dat}, {\tt MHplus\_GammaTot.dat}, {\tt CP\_values.dat},\\*
{\tt BR\_H\_NP.dat}, {\tt BR\_H\_OP.dat}, {\tt BR\_Hplus.dat},\\
{\tt LEP\_HZ\_CS\_ratios.dat},  {\tt LEP\_H\_ff\_CS\_ratios.dat},
{\tt LEP\_2H\_CS\_ratios.dat},\\ 
{\tt LEP\_HpHm\_CS\_ratios.dat}.
\end{quote}

As a third example, if {\tt whichinput = 'hadr'}, {\tt whichanalyses = 'onlyT'}, {\tt nHzero > 0} and {\tt nHplus > 0} 
 {\tt HiggsBounds} requires the input files: 
\begin{quote}
{\tt MH\_GammaTot.dat}, {\tt MHplus\_GammaTot.dat}, {\tt CP\_values.dat},\\* 
{\tt BR\_H\_NP.dat}, {\tt BR\_H\_OP.dat}, {\tt BR\_t.dat}, {\tt BR\_Hplus.dat},\\* 
{\tt TEV\_1H\_hadCS\_ratios.dat} .
\end{quote}

In each of these three examples, {\tt HiggsBounds} 
will also read the file 
\begin{quote}
{\tt additional.dat}
\end{quote}
if it exists.  

As for the subroutine version, if the user does not require processes
involving a particular branching ratio or cross section ratio to be
checked by {\tt HiggsBounds}, that particular array can be filled with
zeros. 
\medskip

\begin{table}
\begin{tabular}{lll}
\hline
{\tt whichinput = 'part'},            &{\tt 'hadr'},  &{\tt 'effC'}\\
\hline
{\tt MH\_GammaTot.dat}                &{\tt MH\_GammaTot.dat}            & {\tt MH\_GammaTot.dat}\\ 
{\tt MHplus\_GammaTot.dat}\mystar     &{\tt MHplus\_GammaTot.dat}\mystar & {\tt MHplus\_GammaTot.dat}\mystar\\
{\tt BR\_H\_NP.dat}\mystar            &{\tt BR\_H\_NP.dat}\mystar        &{\tt effC.dat}\mystarbrac  \\
{\tt BR\_H\_OP.dat}\mystar            & {\tt BR\_H\_OP.dat}\mystar       & {\tt BR\_H\_NP.dat}\mystar \\  
{\tt BR\_t.dat}\mystar                &{\tt BR\_t.dat}\mystar            &{\tt BR\_t.dat}\mystar\\  
{\tt BR\_Hplus.dat}\mystar            &{\tt BR\_Hplus.dat}\mystar        &{\tt BR\_Hplus.dat}\mystar\\
{\tt LEP\_HZ\_CS\_ratios.dat}         & {\tt LEP\_HZ\_CS\_ratios.dat}    & {\tt LEP\_HpHm\_CS\_ratios.dat}\mystar\\
{\tt LEP\_H\_ff\_CS\_ratios.dat}\mystar &{\tt LEP\_H\_ff\_CS\_ratios.dat}\mystar  & {\tt additional.dat}\\
{\tt LEP\_2H\_CS\_ratios.dat}         &{\tt LEP\_2H\_CS\_ratios.dat}     & \\
{\tt LEP\_HpHm\_CS\_ratios.dat}\mystar  &{\tt LEP\_HpHm\_CS\_ratios.dat}\mystar   &\\
{\tt TEV\_H\_0jet\_partCS\_ratios.dat}& {\tt TEV\_1H\_hadCS\_ratios.dat}\mystarbrac & \\   
{\tt TEV\_H\_1jet\_partCS\_ratios.dat}&  {\tt CP\_values.dat}\mystar            &\\  
{\tt TEV\_HW\_partCS\_ratios.dat}     &{\tt additional.dat} &\\  
{\tt TEV\_HZ\_partCS\_ratios.dat}     &&\\  
{\tt TEV\_H\_vbf\_hadCS\_ratios.dat}  && \\
{\tt TEV\_H\_tt\_hadCS\_ratios.dat}\mystar  && \\
{\tt CP\_values.dat}\mystar           && \\
{\tt additional.dat}                  && \\
\hline
\end{tabular}\\
\caption{\sl\label{table:instructions5}The list of possible input files for each
  value of {\tt whichinput}. Note that some input files may not be relevant,
  depending on the values of {\tt whichanalyses}, {\tt nHzero} and {\tt
  nHplus}.  In this case, they are not required.  See
  \refta{table:instructions6} and \refta{table:instructions6a} for more
  details.
\vspace*{5mm}} 
\end{table}

\begin{table}
\begin{tabular}{llllll}
\hline
name of input file & \multicolumn{5}{l}{values of {\tt whichanalyses}}     \\
     & \multicolumn{5}{l}{which this file is relevant to}  \\
\hline
                                       &LandT & onlyL& onlyT& singH   & onlyP\mystar\\
\hline
{\tt MH\_GammaTot.dat}                 &y     & y     & y     & y     &y       \\
{\tt MHplus\_GammaTot.dat}\mystar      &y     & y     & y     & y     &y       \\
{\tt effC.dat}\mystarbrac		       &y     & y     & y     & y     &y       \\   
{\tt LEP\_HZ\_CS\_ratios.dat}          &y     & y     &       & y     &y      \\  
{\tt LEP\_H\_ff\_CS\_ratios.dat}\mystar&y     & y     &       & y     &y       \\  
{\tt LEP\_2H\_CS\_ratios.dat}          &y     & y     &       &       &y       \\
{\tt LEP\_HpHm\_CS\_ratios.dat}\mystar &y     & y     &       & y     &y      \\
{\tt TEV\_H\_0jet\_partCS\_ratios.dat} &y     &       & y     & y     &y      \\   
{\tt TEV\_H\_1jet\_partCS\_ratios.dat} &y     &       & y     & y     &y      \\  
{\tt TEV\_HW\_partCS\_ratios.dat}      &y     &       & y     & y     &y      \\  
{\tt TEV\_HZ\_partCS\_ratios.dat}      &y     &       & y     & y     &y      \\ 
{\tt TEV\_H\_vbf\_hadCS\_ratios.dat}   &y     &       & y     & y     &y      \\ 
{\tt TEV\_H\_tt\_hadCS\_ratios.dat}\mystar&y     &       & y     & y     &y      \\
{\tt TEV\_1H\_hadCS\_ratios.dat}\mystarbrac &y     &       & y     & y     &y      \\ 
{\tt BR\_H\_OP.dat}\mystar              &y     & y     & y     & y     &y      \\   
{\tt BR\_H\_NP.dat}\mystar              &y     & y     & y     & y     &y     \\
{\tt BR\_t.dat}\mystar			&y     &       & y     & y     &y     \\
{\tt BR\_Hplus.dat}\mystar		&y     & y     & y     & y     &y    \\
{\tt CP\_values.dat}\mystar		&y     & y     & y     & y     &y    \\
{\tt additional.dat}  (optional)       &y     & y     & y     & y     &y   \\
\hline
\end{tabular}\\
\caption{\sl\label{table:instructions6}List of input files, specifying which
  values of {\tt whichanalyses} each input file is relevant to (marked by
  'y').  Note that some input files may not be relevant, depending on the
  values of {\tt whichinput}, {\tt nHzero} and {\tt nHplus}.  In this case,
  they are not required.  See \refta{table:instructions5} and
  \refta{table:instructions6a} for more details.
\vspace*{3mm}} 
\end{table}

\begin{table}
\begin{tabular}{lll}
\hline
name of input file & only relevant if & only relevant if\\
                   & {\tt nHzero > 0}     &  {\tt nHplus\mystar\ > 0}\\
\hline
{\tt MH\_GammaTot.dat}                 & y   &       \\
{\tt MHplus\_GammaTot.dat}\mystar             &    &   y    \\
{\tt effC.dat}\mystarbrac                         & y   &       \\
{\tt LEP\_HZ\_CS\_ratios.dat}          & y   &       \\ 
{\tt LEP\_H\_ff\_CS\_ratios.dat}\mystar       & y   &       \\ 
{\tt LEP\_2H\_CS\_ratios.dat}          & y   &       \\
{\tt LEP\_HpHm\_CS\_ratios.dat}\mystar        &    &   y    \\
{\tt TEV\_H\_0jet\_partCS\_ratios.dat} & y   &       \\
{\tt TEV\_H\_1jet\_partCS\_ratios.dat} & y   &       \\
{\tt TEV\_HW\_partCS\_ratios.dat}      & y   &       \\
{\tt TEV\_HZ\_partCS\_ratios.dat}      & y   &       \\
{\tt TEV\_H\_vbf\_hadCS\_ratios.dat}   & y   &       \\
{\tt TEV\_H\_tt\_hadCS\_ratios.dat}\mystar    & y   &       \\
{\tt TEV\_1H\_hadCS\_ratios.dat}\mystarbrac       & y   &       \\
{\tt BR\_H\_OP.dat}\mystar                    & y   &       \\ 
{\tt BR\_H\_NP.dat}\mystar                    & y   &       \\
{\tt BR\_t.dat}\mystar                        &    &   y    \\
{\tt BR\_Hplus.dat}\mystar                    &    &   y    \\
{\tt CP\_values.dat}\mystar                   & y  &        \\
{\tt additional.dat}                   & (optional)   & (optional)      \\
\hline
\end{tabular}
\caption{\sl\label{table:instructions6a} List of files, showing which relate
  to the neutral Higgs searches and which relate to the charged Higgs
  searches. Note that some input files may not be relevant, depending on
  the values of {\tt whichinput} and {\tt whichanalyses}.  In this case,
  they are not required.  See \refta{table:instructions5} and
  \refta{table:instructions6} for more details. }
\end{table}

\subsubsection*{Output file format}

When the command line version of {\tt HiggsBounds} is used, the output
is written to the file {\tt
  <prefix>HiggsBounds\_results.dat}. A sample of the output is shown in
\reffi{fig:sampleoutput}. The key to the process numbering is written
to {\tt <prefix>Key.dat}.

\begin{figure}[ht]
{\scriptsize
\begin{verbatim}
 # generated with HiggsBounds on 08.02.2011 at 11:18
 # settings: LandT, effC
 #
 # column abbreviations
 #   n          : line id of input
 #   Mh(i)      : masses of neutral Higgs boson
 #   Mhplus(i)  : masses of singly, positively charged Higgs boson masses
 #   HBresult   : scenario allowed flag (1: allowed, 0: excluded, -1: unphysical)
 #   chan       : most sensitive channel (see below). chan=0 if no channel applies
 #   obsratio   : ratio [sig x BR]_model/[sig x BR]_limit (<1: allowed, >1: excluded)
 #   ncomb      : number of Higgs bosons combined in most sensitive channel
 #   additional : optional additional data stored in <prefix>additional.dat (e.g. tan beta)
 #
 # channel numbers used in this file
 #           3 : (ee)->(h3)Z->(b b)Z   (LEP table 14b)
 #           4 : (ee)->(h1)Z->(tau tau)Z   (LEP table 14c)
 #         124 : (pp)->W(h1)->l nu (b b)   (CDF Note 9463)
 #         134 : (pp)->h2->tau tau   (arXiv:0805.2491)
 #         157 : (pp)->h1+... where h1 is SM-like  (arXiv:0804.3423 [hep-ex])
 # (for full list of processes, see Key.dat)
 #
 #cols: n    Mh(1)     Mh(2)     Mh(3)   Mhplus(1) HBresult  chan    obsratio     ncomb  additional(1)
 #
        1   359.121   271.963   134.929   100.000        1   134    0.212206E-03     1    0.246862   
        2   75.0123   92.8677   71.9716   100.000        1     4    0.306172E-01     1    0.714964   
        3   136.293   345.483   330.026   100.000        1   124    0.640713E-01     1    0.434594   
        4   111.377   220.765   51.7469   100.000        1     3    0.162811         1    0.727173   
        5   186.131   355.002   146.448   100.000        0   157     15.2354         1    0.230522   
\end{verbatim}
}
\caption{\sl\label{fig:sampleoutput} Sample output file (written to {\tt
    <prefix>HiggsBounds\_Results.dat})} 
\end{figure}

\subsubsection*{Examples}

The {\tt HiggsBounds} package includes a full set of sample input files for
the case $n_H=3$, $n_{H^+}=1$, contained in the folder {\tt example\_data}. 
Each filename is
prefixed with {\tt HB\_randomtest50points\_}. 
To run the command-line version of 
{\tt HiggsBounds} with these files as input, use, for example,
\begin{verbatim}
./configure
make
./HiggsBounds LandT effC 3 1 'example_data/HB_randomtest50points_'
\end{verbatim}
where the values of {\tt whichanalyses} and {\tt whichinput} can be varied
as desired.
The setting ${\tt nHplus}=0$ can be used if
the user does not wish 
to test the charged Higgs sector.
E.g.
\begin{verbatim}
./HiggsBounds LandT effC 3 0 'example_data/HB_randomtest50points_'
\end{verbatim}

\phantom{dddddddd} 

\phantom{dddddddd} 

\phantom{dddddddd} 

\phantom{dddddddd} 

\phantom{dddddddd} 

\phantom{dddddddd} 

\phantom{dddddddd} 

\phantom{dddddddd}

\phantom{dddddddd} 

\phantom{dddddddd} 

\phantom{dddddddd} 

\phantom{dddddddd}

\subsection{Online version}

The online version can be reached via the website%
\footnote{The former website {\tt www.ippp.dur.ac.uk/HiggsBounds} should
redirect to the new one.}
\begin{quote}
{\tt http://projects.hepforge.org/higgsbounds/} .
\end{quote}
It allows the user to select the required number of neutral Higgs
and charged Higgs bosons and then generates a {\tt html} form accordingly. 
The values of {\tt whichinput} and {\tt whichanalyses} can be chosen and the
appropriate theoretical input entered.  {\tt Higgs\-Bounds} will then be
called with these settings and the result outputted to screen.  The online
version contains the additional feature that it notifies the user about the
processes with the second and third highest statistical sensitivities and
the values of {\tt obsratio} for these processes.  This is designed to give
guidance to the user who, for example, wishes to find an excluded region
iteratively by adjusting the input quantities.

The website also contains a selection of pre-filled {\tt html} forms as
examples, including entries for the SM, the fermiophobic Higgs model and the
MSSM with real and complex parameters.

\mystar\ 
There is also the possibility of copying and pasting the results from the
online version of {\tt FeynHiggs} 
(the `FeynHiggs User Control Center' \cite{FH-website}) into a
box on the {\tt Higgs\-Bounds} website for the CP-conserving MSSM.
This text
is then converted to a filled-in {\tt Higgs\-Bounds} input {\tt html} form
(which can be further edited by the user if necessary) that can then be
immediately submitted to {\tt HiggsBounds}.

\section{Physics Applications}
\label{sec:applications}

\subsection{Model scenarios with invisible Higgs decay modes}

The decay of a Higgs boson into stable, weakly interacting neutral particles
(invisible Higgs decay) is allowed in a wide variety of models.  In the
R-parity conserving MSSM with a neutralino $\chi^0_1$ as the lightest
supersymmetric particle (LSP), the decay $h^0 \to \chi^0_1\chi^0_1$ is
dominant in some scenarios~\cite{Griest:1987qv,hep-ph/9603368}.  Invisible
Higgs decays can also be important in models with hidden sectors that couple
to the Higgs sector~\cite{hep-ph/9608245}, models where a Higgs boson can
decay into a pair of
Majorons~\cite{hep-ph/9601269,Chikashige:1980ui,Joshipura:1992ua}, models
with 4th generation neutrinos~\cite{hep-ph/0210153}, non-linear
supersymmetric models in which the Higgs boson can decay into a Goldstino
and a neutralino~\cite{hep-ph/0410165} and some extra dimension
models~\cite{hep-ph/0002178,hep-ph/9903259,hep-ph/0404056,arXiv:0902.1512,hep-ph/9804398,hep-ph/9803315}.

We consider here a simple model in which we vary the branching ratio of the
invisible Higgs decay mode, while keeping all other cross sections and decay
widths fixed to SM values and assuming that the narrow width
approximation holds.  This is achieved with the following program:

{\scriptsize
\begin{verbatim}
!******************************************************
program example_invisible
!******************************************************
 implicit none
 integer :: nH,nHplus               ! HB input
 integer :: HBresult,chan,ncombined ! HB output
 integer :: p,q,pend,qend           ! used in do loops
 double precision :: obsratio       ! HB output
 double precision :: mass           ! Higgs mass
 double precision :: SMGammaTotal   ! SM total decay width for a SM Higgs at this mass 
 ! HB input:
 double precision :: Mh,GammaTotal_hj,                      &  
     &          g2hjss_s,g2hjss_p,g2hjcc_s,g2hjcc_p,        &
     &          g2hjbb_s,g2hjbb_p,g2hjtoptop_s,g2hjtoptop_p,&
     &          g2hjtautau_s,g2hjtautau_p,                  &
     &          g2hjWW,g2hjZZ,g2hjZga,                      &
     &          g2hjgaga,g2hjgg,g2hjhiZ_nHbynH,             &
     &          BR_hjinvisible,BR_hjhihi_nHbynH
 ! HB functions:
 double precision ::SMGamma_h,SMBR_Hbb

 nH=1 ! only one neutral Higgs in model
 nHplus=0 ! no charged Higgs in model
 call initialize_HiggsBounds(nH,nHplus,'LandT')
 pend=201
 qend=201

 open(10,file='example-invisible-results.dat')

 do p=1,pend
   Mh= 50.0D0 + dble(p-1)*200.0D0/dble(pend-1)
   do q=1,qend
    BR_hjinvisible= dble(q-1)*1.0D0/dble(qend-1)

    SMGammaTotal  = SMGamma_h(Mh)
    if(SMGammaTotal.lt.0.0D0)stop'negative result from SMGamma_h(Mh)'

    if(abs(1.0D0-BR_hjinvisible).gt.0.0D0)then
       GammaTotal_hj = SMGammaTotal/(1.0D0-BR_hjinvisible)
    else
       GammaTotal_hj = 1.0D8 !irrelevant if higgs->invisible is the only decay mode
    endif

    ! set normalised couplings to SM values i.e.
    ! all except pseudoscalar couplings set to 1
    ! pseudoscalar couplings set to 0
    g2hjss_s     = 1.0D0;   g2hjss_p     = 0.0D0
    g2hjcc_s     = 1.0D0;   g2hjcc_p     = 0.0D0
    g2hjbb_s     = 1.0D0;   g2hjbb_p     = 0.0D0
    g2hjtoptop_s = 1.0D0;   g2hjtoptop_p = 0.0D0
    g2hjtautau_s = 1.0D0;   g2hjtautau_p = 0.0D0                         

    g2hjWW  =1.0D0
    g2hjZZ  =1.0D0             
    g2hjZga =1.0D0
    g2hjgaga=1.0D0
    g2hjgg  =1.0D0
    g2hjhiZ_nHbynH=0.0D0

    ! only one Higgs in model, so no Higgs cascade decay branching ratio
    BR_hjhihi_nHbynH=0.0D0

    call HiggsBounds_neutral_input_effC( Mh,GammaTotal_hj,  &  
     &          g2hjss_s,g2hjss_p,g2hjcc_s,g2hjcc_p,        &
     &          g2hjbb_s,g2hjbb_p,g2hjtoptop_s,g2hjtoptop_p,&
     &          g2hjtautau_s,g2hjtautau_p,                  &
     &          g2hjWW,g2hjZZ,g2hjZga,                      &
     &          g2hjgaga,g2hjgg,g2hjhiZ_nHbynH,             &
     &          BR_hjinvisible,BR_hjhihi_nHbynH             )

    call run_HiggsBounds( HBresult, chan, obsratio, ncombined )

    write(10,*) Mh,BR_hjinvisible,HBresult,chan,obsratio

   enddo
 enddo
 call finish_HiggsBounds
 close(10)

end program example_invisible
\end{verbatim}
}

which can be compiled using, for example, 

{\scriptsize
\begin{verbatim}
gfortran ./example_invisible.f90 -o ./example_invisible.exe -L<HBdirectorypath> -lHB
\end{verbatim}
}

\subsubsection{Exclusion}

\begin{figure}
\includegraphics[width=0.5\linewidth]{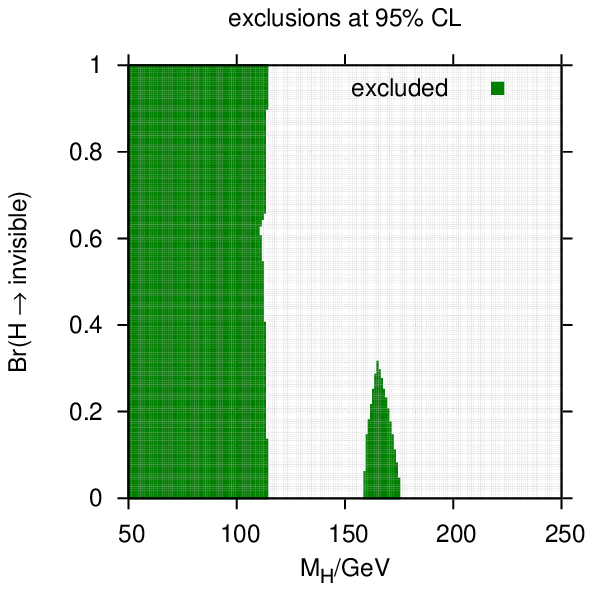}
\includegraphics[width=0.5\linewidth]{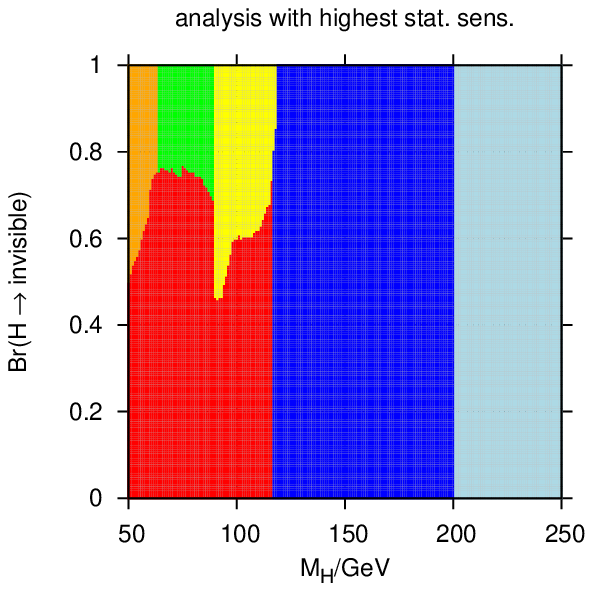}
    \caption{\sl
	\label{fig:invisible} 
Parameter space for our toy model, in which we vary the branching ratio of
the Higgs decay into invisible particles and the Higgs mass.  Higgs production cross
sections and all other Higgs decay widths take SM values.  Left: parameter
points excluded at 95\% C.L. by Higgs searches at LEP and the Tevatron.  Green
(dark grey) = excluded, White = unexcluded.\newline
Right: Channel with the highest statistical sensitivity at each parameter point.
\newline
${\color{gnuplotred}\blacksquare}=e^-e^+\to H Z\to b \bar b Z $, \citere{hep-ex/0602042} (LHWG)\newline
${\color{gnuplotorange}\blacksquare}=e^-e^+\to H Z\to \mathrm{(invisible)} Z $, \citere{arXiv:0707.0373}
(OPAL)\newline
${\color{gnuplotgreen}\blacksquare}=e^-e^+\to H Z\to \mathrm{(invisible)} Z $, \citere{hep-ex/0501033} (L3) \newline
 ${\color{gnuplotyellow}\blacksquare}=e^-e^+\to H Z\to \mathrm{(invisible)} Z $, \citere{hep-ex/0107032} (LHWG) \newline
${\color{gnuplotlightblue}\blacksquare}= p \bar{p} \to H \to W^+ W^-$,  \citere{arXiv:1005.3216} (TEVNPHWG)\newline
${\color{gnuplotblue}\blacksquare}= p \bar{p} \to H+...\to ...$ where $H$ is SM-like, \citere{arXiv:1007.4587} (TEVNPHWG) 
       }
\end{figure}

We show the results from 
the code displayed above
in \reffi{fig:invisible}. In the left
panel, we can see that all values of the invisible Higgs branching ratio are
excluded almost up to the kinematical limit of LEP.  In the right panel, we
can see that LEP Higgsstrahlung processes have the highest statistical
sensitivity in this region, and therefore were used to obtain this
exclusion.  For lower values of invisible Higgs branching ratio, the decay
mode $H\to b \bar b$ was used, whereas at higher values of
BR($H\to$invisible), Higgsstrahlung topologies which explicitly involve the
$H\to$invisible decay mode (i.e.  searches for a $Z$-boson plus large
missing momentum) were used.  The LEP Higgs Working Group combined invisible
Higgs analysis~\cite{hep-ex/0107032} does not cover the region $M_H<90\gev$,
and therefore the OPAL~\cite{arXiv:0707.0373} and L3~\cite{hep-ex/0501033}
analyses are also required.

We can also apply the results from the Tevatron SM Higgs searches directly
to this model, since all the relevant cross sections and branching ratios
differ from their SM equivalents by a common factor (i.e.  the parameter
points all pass the SM-likeness test for the Tevatron SM Higgs analyses, as
described in \refse{sec:listofanalyses}).  Therefore, the exclusion at 95\%
C.L. of a SM Higgs boson between $158\gev$ and $175\gev$ provided by
\citere{arXiv:1007.4587} translates to a wedge-shaped excluded region in the
parameter space of our toy model, extending up to BR($H\to$invisible)=0.32.

\subsubsection{Interpreting the variable {\tt obsratio}}

We can also obtain interesting information about our model from the variable
{\tt obsratio} outputted by the program {\tt example\_invisible}.  Recall
that {\tt obsratio} is the ratio of the theoretical cross section to the
observed cross section for the channel with the highest statistical
sensitivity at each parameter point 
$Q_\MOD(X_0)/Q_\OBS(X_0)$, 
as discussed in \refse{sec:usage}. 
Roughly speaking, this variable gives an idea of `how strong' the
exclusion of a particular
parameter point is.  We shall now demonstrate two possible ways in which
this variable can be used to obtain additional insights into our toy model. 
We will first discuss the excluded low mass region 
(with $\text{\tt obsratio} > 1$)
and then we will discuss the unexcluded intermediate mass region 
(with $\text{\tt obsratio} < 1$).

\begin{figure}[t]
\includegraphics{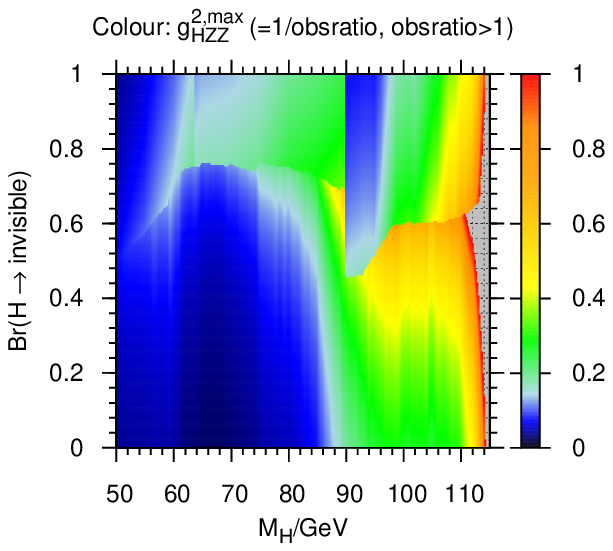}
\includegraphics{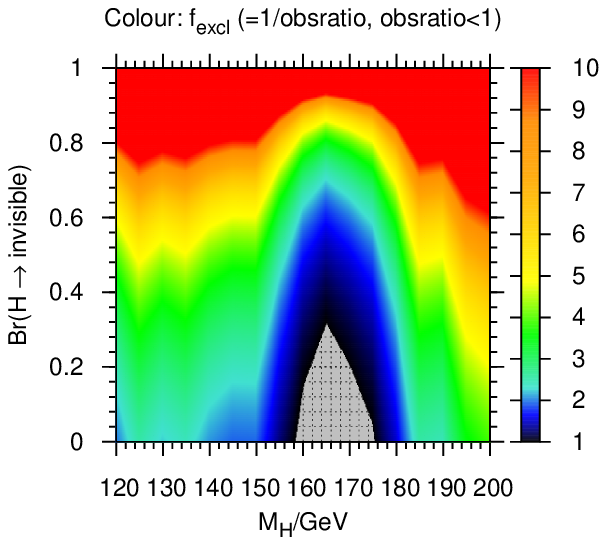}
    \caption{\sl
	\label{fig:invisibleobsratio} 
Subsections of the parameter space for our toy model, in which we vary the
branching ratio of Higgs into invisible particles and the Higgs mass.  {\tt
obsratio} is the ratio of the theoretical cross section to the observed
cross section for the channel with the highest statistical sensitivity, as
calculated by the program {\tt example\_invisible}, plotted when 
$\text{\tt obsratio}>1$ (left) or when $\text{\tt obsratio}<1$ (right).  
Left: we interpret
1/{\tt obsratio} as the maximum value of $g^2_{HZZ}$ which is consistent
with the parameter point being excluded at 95\% C.L.  Right: we interpret
1/{\tt obsratio} as the factor $f_{\rm excl}$ by which the measured SM cross
section limit must be reduced in order to exclude this parameter point at
95\% C.L. (in this plot, red denotes $f_{\rm excl}\geq 10$).}
\end{figure}

In models with extended Higgs sectors, such as the MSSM, the Higgs couplings
to $Z$-bosons frequently obey the sum rule
$\sum_{i=1}^{n_{H^0}}g^2_{h_iZZ}=1$, and therefore the $h_i$-$Z$-$Z$
couplings in these models are suppressed with respect to the SM coupling. 
It is therefore interesting to consider the extent to which the parameter
space of our toy model can still be excluded if we suppress the $H$-$Z$-$Z$
coupling.  We can interpret the inverse of the {\tt obsratio} calculated by
the program {\tt example\_invisible} as the maximum $g^2_{HZZ}$ which will
still allow the parameter point to be excluded at 95\% by the LEP Higgs
search results (considering Higgsstrahlung processes only).  1/{\tt
obsratio} is plotted in \reffi{fig:invisibleobsratio}(left) in the low Higgs
mass region, for $\text{\tt obsratio}>1$.  In this plot, we can see that, for
example, a Higgs boson with $M_H= 95\gev$ and BR($H\to$invisible)=0.5 (and
all other decay widths as in SM) can not be excluded by the LEP
Higgsstrahlung searches if $g^2_{HZZ}<0.5$.

We shall now look at 1/{\tt obsratio} in the unexcluded part of the
intermediate Higgs mass range, where the Tevatron SM Higgs analysis
\citere{arXiv:1007.4587} has the highest statistical sensitivity.  This is
shown in \reffi{fig:invisibleobsratio}(right), 
for $\text{\tt obsratio}<1$.  In
this mass range, it is interesting to interpret 1/{\tt obsratio} as the
factor $f_{\rm excl}$ by which the current SM cross section limit must be
lowered
 in order to obtain a 95\% exclusion in our toy model.  Therefore,
if, for example, a future Tevatron analysis can 
lower
the current limit by
10\% at $M_H=165\gev$, this will allow BR($H\to$invisible) up to 0.38 to be
excluded in our toy model at this Higgs mass.  In this way, we can get an
impression of which areas of parameter space are likely to be testable in
the near future and which areas will require much more data.

\subsection{Constraints on the Randall-Sundrum scalar sector}

The Randall-Sundrum (RS) Model, RS1, considers spacetime a slice of 5d
anti-de-Sitter space with two boundaries, the IR brane
(our 4d spacetime) and the UV brane \cite{RS}. This configuration 
yields an explanation for the hierarchy problem, 
i.e. why the mass scale of electroweak physics is so much smaller than 
the Planck scale $M_\text{Pl}$. 
The spacetime metric can be written as 
$$
ds^2 = e^{-2 k r_c y}\eta_{\mu\nu}dx^\mu dx^\nu - r_c^2 dy^2\,,
\;\;y\in [0,\pi]\,,
$$
with $x^\mu$ ($\mu\in{0,1,2,3}$) coordinates of the usual 4d spacetime 
and $y$ the coordinate of the fourth spatial direction.
The quantities $k$, $r_c^{-1}$ are ${\cal O}(M_\text{Pl})$ 
with $k r_c \approx 12$.
The ``little hierarchy'' between $k$ and $r_c^{-1}$ can be generated and 
stabilized naturally \cite{GWa}.
The hierarchy between electroweak and Planck scale is a consequence
of the ``warped'' metric: mass parameters in the fundamental 5d model $m_0$
appear in our visible space as 
$$
m = m_0 e^{-k r_c \pi} \approx m_0 10^{-16}\,.
$$
Hence, all mass scales of the model can be assumed to be of the order of the Planck
scale.
 Many modifications to the original RS Model,
 concerning  which fields should be localised 
 on the branes, have been considered necessary in order to meet
 electroweak precision and flavour constraints
 \cite{Neubert-etal-and-refs-therein}.
 However, the Higgs field needs to be localised on or near
 the IR brane in order to explain the hierarchy problem. 
 Therefore, the RS scalar sector is a rather robust prediction
 of RS models.

\subsubsection{The Higgs--radion system} 

Fluctuations of the 
compactification ``radius'' $r_c$ correspond to a
scalar degree of freedom in addition to the 4d metric fluctuations.
As a consequence of stabilising $r_c$, 
this scalar acquires a vacuum expectation value (VEV) $\Lambda_\phi$
and appears as a massive scalar (the radion $\phi_0(x)$)
in the spectrum \cite{GWa,GWb}. 
Higgs--radion mixing may occur via the curvature--scalar interaction
\cite{xiHsquared-term}
$$
{\cal L}  = -\xi \sqrt{-g_{\text{ind}}}\: R(g_{\text{ind}})\: \Phi^\dagger \Phi
\,,
$$
where $g_{\text{ind}}(\phi_0(x),\cdots)$ is the induced 4d metric on the IR brane,
$R$ the Ricci scalar, and $\Phi$ the Higgs doublet.
Hence, in general, what would be for $\xi=0$
the physical Higgs degree of freedom {$h_0$} 
and the radion {$\phi_0$}
mix to form two mass eigenstates $h$ and $\phi$.
One arrives at the fields $h_0$ and $\phi_0$
after rescaling to canonical normalisation on the IR brane.
The part of the Lagrangian bilinear in $h_0$ and $\phi_0$
reads \cite{CsakiGraesserKribs,Gunion-et-al}:
\begin{align}
{\cal L} & = -\frac{1}{2}(1+6\gamma^2\xi)\phi_0\square\phi_0
	-\frac{1}{2} h_0\square h_0
	-6\gamma\xi\phi_0\square h_0
	-\frac{1}{2}m_{\phi_0}^2 \phi_0^2
	-\frac{1}{2}m_{h_0}^2 h_0^2
\,.
\end{align}
Diagonalisation and canonical normalisation of the bilinear 
terms can be obtained for the physical fields $h$ and $\phi$ via
the identification:
\begin{align}
\label{abcd}
\phi_0 & = a \phi + b h\,, &  h_0 & = d h + c \phi\,,
\end{align}
with 
\begin{align}
a&=-\frac{\cos\theta}{Z} \,, 
	& b&=\frac{\sin\theta}{Z} \,,
&
c&= \sin\theta +\frac{6\xi\gamma}{Z}\cos\theta\,,
	& d&= \cos\theta-\frac{6\xi\gamma}{Z}\sin\theta\,, 
\end{align}
and
\begin{align}
 \tan 2\theta & = \frac{12\gamma\xi Z m_{h_0}^2}
		   	{m_{\phi_0}^2 - m_{h_0}^2(Z^2-36\xi^2\gamma^2)}\,,
& Z^2 &= 1 + 6\xi\gamma^2(1-6\xi)\,, & \gamma &= \frac{v}{\Lambda_\phi}\,.
\end{align}
In this formulae, $v$ ($\approx 246$ GeV) is the Higgs VEV.
The radion VEV $\Lambda_\phi$ is a free parameter of the model.
The masses and couplings to known matter and gauge bosons
of the Higgs--radion sector are determined
by choosing values 
for the four parameters $\Lambda_\phi$, 
the curvature--Higgs coupling parameter $\xi$, and the masses 
$m_h$ and $m_\phi$ of the physical fields.
However, those parameter choices are subject to theoretical constraints,
which come from requiring positive kinetic energy and mass terms
in the Lagrangian \cite{CsakiGraesserKribs,Gunion-et-al}:
\begin{align}
\label{th constraints}
Z^2 &  > 0 \,,
& 
\max\left[\frac{m_h^2}{m_\phi^2} , \frac{m_\phi^2}{m_h^2}\right]
	& >  1 + \frac{2}{x} \left(\sqrt{1-x} + {1-x}\right)
\,,
\end{align} 
with 
$x=(1+36\xi^2\gamma^2/Z^2)^{-1}$. 
The Lagrangian parameters $m_{h^0}^2$ and $m_{\phi^0}^2$ are determined for 
a given valid parameter scenario by
\begin{align*} 
 m_{h^0}^2 & = \frac{x}{2}(\Sigma +[-] \Delta)\,,\; 
	m_{\phi^0}^2 = \frac{Z^2}{2}(\Sigma -[+] \Delta)\,, 
	\text{ if $m_h > m_\phi$ $[m_h < m_\phi]$}\,,\\
 & \text{with } 
 \Sigma  = m_h^2 + m_\phi^2\,,\;
 \Delta = \sqrt{\Sigma^2 - \frac{4}{x} m_h^2 m_\phi^2}
\,.
\end{align*} 
In the following, we do not consider scenarios where the physical scalars 
can decay into Kaluza--Klein (KK) excitations of the graviton, i.e. implicitly
we assume that the mass $m_1$ of this excitation is large enough
to forbid such decay channels.
This puts a lower bound on the curvature parameter $k$.

Providing values for $\Lambda_\phi$, $\xi$, $m_h$ and $m_\phi$
for a valid scenario, the quantities $a$, $b$, $c$, and $d$ 
defined in Eq.~(\ref{abcd}) are determined.
This, in turn, determines
the effective couplings of the two physical scalars $h$ and $\phi$ to
gauge bosons and matter normalised to the couplings of a SM Higgs boson
with the same mass \cite{hep-ph/0002178,Gunion-et-al}:
\begin{align*}
\frac{\Gamma(h\to f\bar f)}{\Gamma(H_\SM\to f\bar f)} 
& = \frac{\Gamma(h\to W^+ W^-)}{\Gamma(H_\SM\to W^+ W^-)} 
  = \frac{\Gamma(h\to ZZ)}{\Gamma(H_\SM\to ZZ)} 
	= \left|d + \frac{v}{\Lambda_\phi} b\right|^2 \,,\\
\frac{\Gamma(\phi\to f\bar f)}{\Gamma(H_\SM\to f\bar f)} 
& = \frac{\Gamma(\phi\to W^+ W^-)}{\Gamma(H_\SM\to W^+ W^-)} 
  = \frac{\Gamma(\phi\to ZZ)}{\Gamma(H_\SM\to ZZ)} 
	= \left|c + \frac{v}{\Lambda_\phi} a\right|^2\,,\\
\frac{\Gamma(h\to gg)}{\Gamma(H_\SM\to gg)} 
	& = \frac{v^2}{\Lambda_\phi^2}
	\frac{|2 b b_3 - (b + \frac{\Lambda_\phi}{v} d)F_{1/2}(\tau^h_t)|^2}{|F_{1/2}(\tau^h_t)|^2}\,,\\
\frac{\Gamma(\phi\to gg)}{\Gamma(H_\SM\to gg)} 
	& = \frac{v^2}{\Lambda_\phi^2}
	\frac{|2 a b_3 - (a + \frac{\Lambda_\phi}{v} c)F_{1/2}(\tau^\phi_t)|^2}{|F_{1/2}(\tau^\phi_t)|^2}
	\,,\\
\frac{\Gamma(h\to \gamma\gamma)}{\Gamma(H_\SM\to \gamma\gamma)} 
	& = \frac{v^2}{\Lambda_\phi^2}
	\frac{|3b(b_2+b_Y) - (b + \frac{\Lambda_\phi}{v} d)
	(4 F_{1/2}(\tau^h_t) + 3F_1(\tau^h_W))|^2}{|4 F_{1/2}(\tau^h_t) + 3F_1(\tau^h_W)|^2}\,,\\
\frac{\Gamma(\phi\to \gamma\gamma)}{\Gamma(H_\SM\to \gamma\gamma)} 
	& = \frac{v^2}{\Lambda_\phi^2}
	\frac{|3a(b_2+b_Y) - (a + \frac{\Lambda_\phi}{v} c)
	(4 F_{1/2}(\tau^\phi_t) + 3F_1(\tau^\phi_W))|^2}{|4 F_{1/2}(\tau^\phi_t) + 3F_1(\tau^\phi_W)|^2}
	\,,
\end{align*}
with $b_2=19/6$, $b_3=7$, $b_Y=-41/6$, $\tau^{h[\phi]}_t = 4 m_t^2/m_{h[\phi]}^2$,
$\tau^{h[\phi]}_W = 4 m_W^2/m_{h[\phi]}^2$, and
\begin{align*}
F_{1/2}(\tau) & = -2\tau(1+(1-\tau)f(\tau))\,, & 
F_1(\tau) & = 2+3\tau+3\tau(2-\tau)f(\tau)\,,\\
& & f(\tau)&=\left\{
\begin{array}{ll}
\arcsin^2(\tau^{-1/2})\,, 
	& \tau \geq 1\\
-\frac{1}{4}\left[\ln\left(
	\frac{1+\sqrt{1-\tau}}{1-\sqrt{1-\tau}}-i\pi \right)\right]
	& \tau > 1
\end{array}
\right.\,.
\end{align*}  
Like for the Higgs, the radion couplings to massive fermions
and gauge bosons are proportional to mass, but e.g. the couplings
$\phi_0\, b \bar b$ and $\phi_0\, \gamma\gamma$ are suppressed
while $\phi_0\, g g$ is enhanced 
with respect to the SM Higgs boson. 
In order to
consider all possible search channels for the Randall-Sundrum model
with \HB,
knowledge of the branching ratios for the decays  
$h \to \phi\phi$ and $\phi \to h h$ is needed as well
\footnote{Formulas for those branching ratios can be obtained from
\cite{Gunion-et-al}.}.
However, those channels turn out not to be 
the most significant ones in the examples we study below.

\subsubsection{Constraints}

We chose here a rather extreme
scenario with $\Lambda_\phi$ as low as 1 TeV, 
the phenomenology of which has been pioneered in \cite{Gunion-et-al}.
This scenario may already have been excluded by other observations.
In particular, our assumption that $m_1$ is high enough such that 
decays of $h$ or $\phi$ into graviton KK excitations are impossible,
entails a rather high curvature scale $k$ in conflict e.g. with
direct searches
for graviton KK excitations in the Randall-Sundrum model
\cite{RSKKgraviton-search}. 
However, it serves demonstration purpose well.
To our knowledge, this is the first time that 
Tevatron Higgs search results have been recast
in order to provide constraints on the scalar
sector of the Randall-Sundrum model.
A 
more thorough 
phenomenological study 
will appear elsewhere \cite{RS-detailed-constraints}.

\begin{figure}[t]
\includegraphics[width=1.0\textwidth]{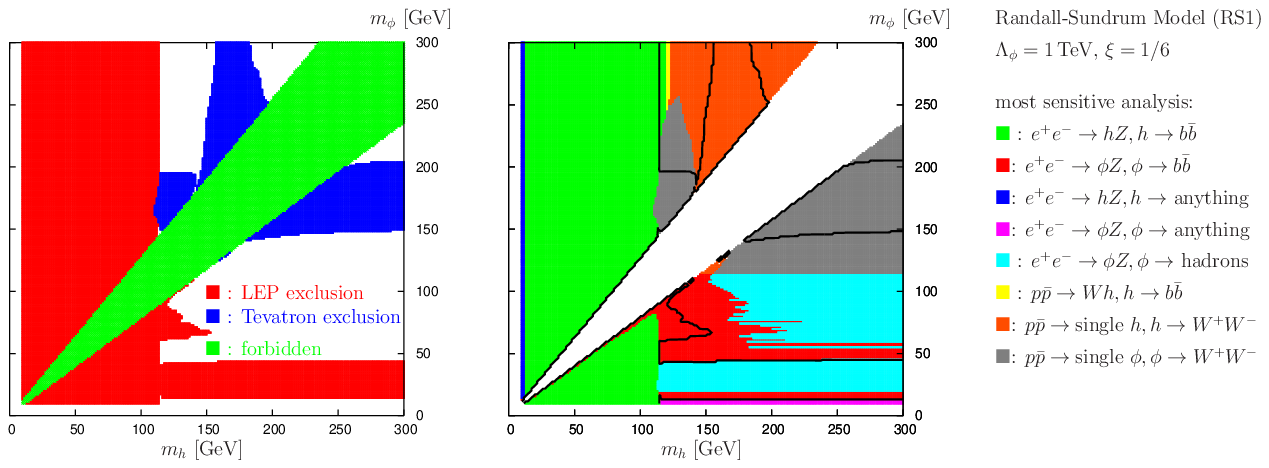}
\caption{\label{fig4} \sl Excluded region in the $m_h-m_\phi$ plane
	of the scalar sector of the original Randall-Sundrum model (RS1) 
 	for $\Lambda_\phi = 1\,\tev$ and $\xi=1/6$ 
	(left panel)
	and most sensitive search channel (right panel)
	using \HB\ \vers{2}{0}{0}.
}
\end{figure}

Fig.~\ref{fig4} shows for the RS model the excluded region (left panel)
and the search channel with highest sensitivity (right panel)
in the $m_h-m_\phi$ plane, while the two other free parameters 
in this model are set to $\Lambda_\phi = 1\,\tev$ and $\xi = 1/6$.
Note the region of parameter space inaccessible due to
the theoretical constraints in Eq.~(\ref{th constraints})
shown in Fig.~\ref{fig4} (left panel).
This scenario shows slight Higgs--radion mixing, i.e. $h$ behaves 
mainly like the SM Higgs and $\phi$ mainly like the unmixed radion. 
This is reflected by the LEP exclusion of $m_h$ values up to
almost the SM mass limit via the process $e^+ e^-\to hZ, h\to b\bar b$
(see Fig.~\ref{fig4}, right panel)
almost independent of $m_\phi$.
For larger values of $m_h$, there is a LEP excluded region
via $e^+ e^-\to \phi Z, \phi\to$ hadrons for $m_\phi$ below about 50 GeV.
In this region, the enhanced $gg$ coupling and the suppressed 
$b\bar b$ coupling of the radion-like $\phi$ renders this channel
the most sensitive one.
The exclusion of parameter regions by Tevatron searches
is mainly due to the model independent cross section limit on 
the process $p\bar p \to \text{single $S$}, S\to W^+W^- \to l\nu l\nu$
\cite{arXiv:1005.3216} for a scalar~$S$. 
Indeed, for scalar masses $m_h$ and $m_\phi$ above around 120 GeV
in the displayed parameter region, this analysis has the 
highest statistical sensitivity (see Fig.~\ref{fig4}, right panel).
Furthermore, the highest exclusion power of this analysis
is for scalar masses around $2 m_W\approx 165\,\gev$.
The right panel of 
Fig.~\ref{fig4} reveals that sizable regions where either $m_h$ or
$m_\phi$ have values close to 165 GeV
are excluded by this analysis with $S=h$ or $S=\phi$.
For instance for $m_h=300\,\gev$, the excluded range of $m_\phi$ values 
due to non-observation of the radion-like scalar $\phi$
is wider than for a SM-like Higgs because of the net enhancement of the  
signal cross section $\sigma(p\bar p \to gg \to \phi) \times \BR(\phi\to W^+W^-)$
with respect to the SM.

\begin{figure}[t]
\includegraphics[width=1.0\textwidth]{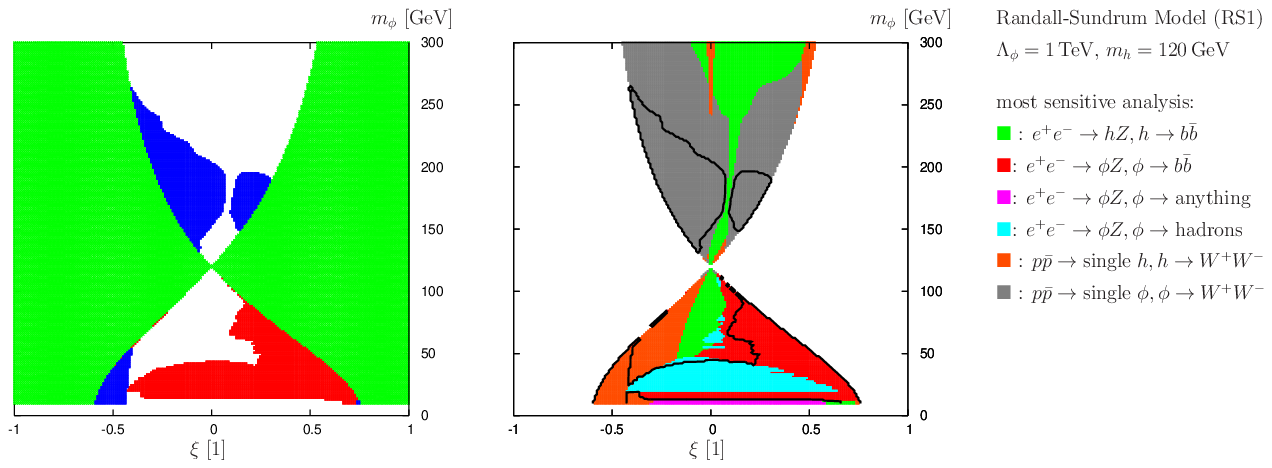}
\caption{\label{fig5} \sl Excluded region in the $\xi-m_\phi$ plane
	of the scalar sector of the original Randall-Sundrum model (RS1) 
 	for $\Lambda_\phi = 1\,\tev$ and $m_h = 120\,\gev$ 
	(left panel, see Fig.~\ref{fig4} (left panel) for the key)
	and most sensitive search channel (right panel)
	using \HB\ \vers{2}{0}{0}.
}
\end{figure}

Fig.~\ref{fig5} shows the excluded Randall-Sundrum parameter space 
also for $\Lambda_\phi = 1\,\tev$ but as function of $\xi$ and $m_\phi$ 
with fixed $m_h = 120\,\gev$.\footnote{
Fig.~\ref{fig5} can be seen as an update of Fig.~18 in \cite{Gunion-et-al}.
}
The parameter exclusion for $m_\phi$ above 120 GeV is entirely 
due to the application of the 
Tevatron analysis \cite{arXiv:1005.3216} to the radion-like $\phi$,
i.e. due to the search channel
$p\bar p \to \text{single $\phi$}, \phi\to W^+W^- \to l\nu l\nu$ 
(see Fig.~\ref{fig5}, right panel).
In this region, the effective $gg\phi$ coupling is rising with 
falling $\xi$, while the effective $WW\phi$ coupling 
goes through a minimum located near $\xi \approx 0$.
This explains, on the one hand, the general trend of an
increasing excluded $m_\phi$-interval with falling $\xi$
values and, on the other hand, the unexcluded funnel region
near $\xi \approx 0$.
For $m_\phi$ below 120 GeV, another region is 
excluded by applying the Tevatron analysis 
\cite{arXiv:1005.3216} --- this time to the Higgs-like $h$.
This analysis application becomes the most sensitive one
because in the lower left part of the allowed parameter space, both
the effective $ggh$ and $WWh$ coupling rise sufficiently with falling 
$\xi$.
The LEP excluded region consists of one region where the most sensitive channel is
$e^+e-\to\phi Z,\phi\to$ hadrons
\cite{hep-ex/0510022,hep-ex/0205055,hep-ex/0312042,hep-ex/0408097} 
(for $m_\phi$ roughly below 50 GeV in Fig.~\ref{fig5}, right panel),
and another one where it is $e^+e-\to\phi Z,\phi\to b\bar b$
\cite{hep-ex/0602042}.

\section{Summary}

We have presented the code {\tt HiggsBounds} \vers{2}{0}{0}
which allows to test neutral and 
charged Higgs sectors of arbitrary models against the current exclusion bounds
from the Higgs searches at LEP and the Tevatron.  
The model predictions which are required as input to the program,
and in particular the three variants of how this information 
can be provided by the user,
are described in detail. 
Version 2.0.0 represents a significant upgrade
of the code since its first release.
As a major extension, the code allows now the predictions 
for (singly) charged Higgs bosons to be 
confronted with LEP and Tevatron searches.
Furthermore,
the newly included analyses contain
LEP searches for neutral Higgs bosons ($H$) decaying invisibly or into 
(non flavour tagged) hadrons as well as decay-mode independent searches,
LEP searches via the production modes $\tau^+\tau^- H$ and $b\bar b H$,
and Tevatron searches via $t\bar t H$.
Many updated Tevatron analyses have replaced previously included ones.
In particular, all 
Tevatron results presented at the 
ICHEP'10 are included in \HB\ \vers{2}{0}{0}.
We presented a full list of the analyses (with references) incorporated 
in the program and the implemented conditions for Higgs bosons
to be met in order to be applicable to some of the analyses.
We provide the complete operating instructions for \HB \vers{2}{0}{0}
with examples, indicating new or extended features with respect
to version 1.0.0 of the code.

We have furthermore presented two new physics applications 
of the program. One application considers a SM-like Higgs boson where 
an extra Higgs decay channel to invisible particles is open.
We find, for instance, that the current Tevatron searches for 
a SM-like Higgs boson yield an exclusion interval around 
$m_H \approx 2 m_W$ even for the case of 
a Higgs boson which decays invisibly
with up to 32\% branching ratio.
As a further application we have investigated the scalar sector of the 
Randall-Sundrum
model. Our results show, to our knowledge, for the first time the impact
of Tevatron Higgs search results on this model.
For the parameter scenario we study here, we find that once the Tevatron
results are taken into account in addition to 
the LEP constraints that had been considered before in
\cite{Gunion-et-al}, large portions of the parameter space are excluded
as a consequence of the model-independent Tevatron Higgs search via the decay 
$H \to W^+ W^-$ \cite{arXiv:1005.3216}.

The functionality of {\tt HiggsBounds} \vers{2}{0}{0} is such that 
upcoming Higgs search limits from the LHC can easily be incorporated. 
The bounds from Higgs searches at LEP, the Tevatron and the LHC itself 
will provide an important source of information for testing model
interpretations of possible Higgs-like signals, which will hopefully
soon emerge from the LHC Higgs searches.

\section*{Acknowledgements}

We would like to thank Ralf Bernhard, Peter Bock, Gabriella Pasztor and 
Yvonne Peters for helpful conversations and input.
The work of P.B. was supported by 
the Helmholtz Young Investigator Grant VH-NG-303.
The work of P.B. and G.W. was supported by
the Collaborative Research Center SFB676 of the DFG.
The work of O.Br. and K.E.W was supported by the Helmholtz Alliance
HA-101 'Physics at the Terascale,'
The work of S.H.\ was partially supported by CICYT (grant FPA 2007--66387
and FPA 2010--22163-C02-01).
This work has been supported in part by
the European Community's Marie-Curie Research Training Network 
under contract MRTN-CT-2006-035505
`Tools and Precision Calculations for Physics Discoveries at Colliders'
(HEPTOOLS)
and under contract MRTN-CT-2006-035657
`Understanding the Electroweak Symmetry
Breaking and the Origin of Mass using the First Data of ATLAS'
(ARTEMIS).
We would like to thank the HepForge development environment, where
the web version of HiggsBounds is currently hosted.

\bibliographystyle{h-elsevier3-newarxivid-leftjust.bst}

\end{document}